\documentclass[structabstract]{aa}  %
\usepackage{graphicx}

\usepackage{xcolor}
\usepackage{soul}

\usepackage{hyperref}
\hypersetup{
    colorlinks = true,
    citecolor ={blue}
}

\usepackage{subfigure}

\newcommand{\logg}{\mbox{$\log g$}\xspace}
\newcommand{\Teff}{\mbox{$T_\mathrm{eff}$}\xspace}

\newcommand{\Msol}{$M_\odot$}
\newcommand{\Rsol}{$R_\odot$}
\usepackage{txfonts}
\usepackage{natbib}
\bibpunct{(}{)}{;}{a}{}{,} % to follow the A&A style
\begin{document}  

\title{Pulsating hydrogen-deficient white dwarfs and pre-white dwarfs observed with {\it TESS}}
\subtitle{VI. Asteroseismology of the GW Vir-type central star of the Planetary Nebula NGC 246}

\author{Leila M. Calcaferro\inst{1,2},
        Paulina Sowicka\inst{3},
        Murat Uzundag\inst{4},
        Alejandro H. C\'orsico\inst{1,2}, 
        S. O. Kepler\inst{5}, 
        Keaton J. Bell\inst{6}, 
        Leandro G. Althaus\inst{1,2}, 
        Gerald Handler\inst{3}, 
        Steven D. Kawaler\inst{7}, \and
        Klaus Werner\inst{8}}  
\offprints{lcalcaferro@fcaglp.unlp.edu.ar} 

\institute{$^1$ Grupo  de Evoluci\'on  Estelar y  Pulsaciones,  Facultad de 
           Ciencias Astron\'omicas  y Geof\'{\i}sicas, Universidad
           Nacional de La Plata, Paseo del Bosque s/n, (1900) La
           Plata, Argentina\\   
           $^{2}$ Instituto de Astrof\'{\i}sica
           La Plata, CONICET-UNLP, Paseo  del Bosque s/n, (1900) La
           Plata,
           Argentina\\ 
           $^{3}$ Nicolaus Copernicus Astronomical Center, Polish Academy of Sciences, ul.  Bartycka 18, PL-00-716, Warszawa, Poland\\
           $^{4}$ Institute of Astronomy, KU Leuven, Celestijnenlaan 200D, B-3001 Leuven, Belgium \\
           $^{5}$ Instituto de F\'{\i}sica, Universidade Federal do Rio Grande do Sul, 91501-970 Porto-Alegre, RS, Brazil\\
           $^{6}$ Department of Physics, Queens College, City University of New York, Flushing, NY-11367, USA\\
           $^{7}$ Department of Physics and Astronomy, Iowa State UniversityAmes, IA 50011, USA\\ 
           $^{8}$ Institut f\"ur Astronomie und Astrophysik, Kepler Center for Astro and Particle Physics, Eberhard Karls Universit\"at, Sand 1, 72076 T\"ubingen, Germany\\ 
+
\email{lcalcaferro@fcaglp.unlp.edu.ar}}
\date{Received, }  

\abstract {Significant advances have been achieved through the latest improvements in the photometric observations accomplished by the recent space missions, substantially boosting the study of pulsating stars via asteroseismology. The {\it TESS} mission has already proven to be of particular relevance for pulsating white dwarf and pre-white dwarf stars.}
{We report a detailed asteroseismic analysis of the pulsating PG 1159 star NGC\,246 (TIC\,3905338), the central star of the planetary nebula NGC 246, based on high-precision photometric data gathered by the {\it TESS} space mission.}
{We reduced {\it TESS} observations of NGC\,246 and performed a detailed asteroseismic analysis using fully evolutionary PG~1159 models computed accounting for the complete prior evolution of their progenitors. We constrained the mass of this star by comparing the measured mean period spacing with the average of the computed period spacings of the models, and also employed the observed individual periods to search for a seismic stellar model.}
 {We extracted $17$ periodicities from the {\it TESS} light curves from the two sectors where NGC\,246 was observed. All the oscillation frequencies are associated with $g$-mode pulsations, with periods spanning from $\sim 1460$ to $\sim 1823\ $s. We found a constant period spacing of $\Delta \Pi= 12.9\ $s, allowing us to deduce that the stellar mass is larger than $\sim 0.87\ M_{\odot}$ if the period spacing is assumed to be associated with $\ell= 1$ modes, and $\sim 0.568\ M_{\odot}$ if it is associated with $\ell= 2$ modes. 
 The less massive models are more consistent with the distance constraint from \textit{Gaia} parallax.
  Although we were not able to find a unique asteroseismic model for this star, the period-to-period fit analyses suggest a high-stellar mass ($\gtrsim 0.74\ M_{\odot}$) when the observed periods are associated with modes with $\ell= 1$ only, and both a high ($\gtrsim 0.74\ M_{\odot}$) and intermediate ($\sim 0.57\ M_{\odot}$) stellar mass when the observed periods are associated with modes with $\ell= 1$ and $2$. 
  }
  {}
 
  \authorrunning{Calcaferro et al.}
  \titlerunning{Asteroseismology of the GW Vir star in NGC 246}
  \maketitle
%----------------------------------------------------------------  
   
\section{Introduction}  
\label{intro}  
GW Vir stars are pulsating PG~1159 stars, that is, pulsating hot hydrogen(H)-deficient white dwarfs (WDs) and pre-WDs with surface layers rich in helium (He), carbon (C), and oxygen (O) \citep{2006PASP..118..183W,2014A&A...564A..53W,2015A&A...584A..19W,2019A&ARv..27....7C,2023arXiv230916537S}. The class of GW Vir stars is often separated into the so-called variable planetary nebula nuclei (PNNVs) that are still surrounded by a nebula, and DOVs\footnote{Even though white dwarf stars of spectral type DO do not pulsate, the term "DOVs" has historically been used as a variable star designation for GW Vir pulsators without a nebula.}, objects that lack a nebula. Among GW Vir stars also are the pulsating Wolf-Rayet central stars of a planetary nebula ([WC]) and early-[WC] ([WCE]) stars, since they share the pulsation properties of pulsating PG 1159 stars  \citep{2007ApJS..171..219Q}. PG 1159 stars are thought to be the evolutionary link between post-asymptotic giant branch (AGB) stars and most of the H-deficient WDs \citep{2005A&A...435..631A,2010A&ARv..18..471A,2006PASP..118..183W,2021ApJ...918L...1S}. The origin of these stars is likely to be in the context of a single-star evolution in a born-again episode induced by a delayed (late or very late) post-AGB He thermal pulse \citep{2001ApJ...554L..71H,2001Ap&SS.275....1B,2005A&A...435..631A, 2006A&A...454..845M}, or binary star evolution in a binary WD merger \citep{2022MNRAS.511L..60M, 2022MNRAS.511L..66W}. GW Vir stars exhibit multiperiodic brightness variations with periods ranging from $300$ to $6000\ $s associated with low-degree ($\ell \leq 2$) nonradial gravity($g$)-mode pulsations excited by the $\kappa$ mechanism due to partial ionization of C and O in the outer layers \citep{1983ApJ...268L..27S, 1984ApJ...281..800S,1991ApJ...383..766S,1996ApJ...462..376B,1996ASPC...96..361S,1997A&A...320..811G,2005A&A...438.1013G,2006A&A...458..259C, 2007ApJS..171..219Q}. DOVs and PNNVs share the same mode driving mechanism, regardless of the absence or presence of a nebula. %.

Remarkably, significant observational efforts have been made in the last years to study pulsating WDs and pre-WDs, among them, the GW Vir stars \citep{2019A&ARv..27....7C}. From dedicated ground-based observations \cite[see, e.g., the works from][]{2002A&A...381..122V,2007A&A...467..237F,2008A&A...477..627C,2014MNRAS.442.2278K,2023arXiv230916537S}, to the unprecedented high-quality data brought about by the advent of space missions, asteroseismology of GW Vir stars has been significantly advanced. In the context of space observations, the {\it Kepler} extended mission \citep[{\it K2},][]{2014PASP..126..398H} and its successor, the Transiting Exoplanet Survey Satellite \citep[{\it TESS},][]{2015JATIS...1a4003R}, have promoted this field even further. The high-sensitivity and continuous observation by {\it TESS} of $\sim 85\%$ of the sky, have made the discovery of an enormous number of pulsating stars possible \citep[see, e.g.,][]{2022ARA&A..60...31K}, making valuable contributions in the field of WDs and pre-WDs \citep[][]{2022BAAA...63...48C}. Particularly relevant for the present paper are the works by \cite{2021A&A...645A.117C}, dedicated to the analysis of a set of already known GW Vir stars, and \cite{2021A&A...655A..27U,2022MNRAS.513.2285U}, where a total of four new members of this type of stars were presented. In addition, combining {\it Kepler/K2} and {\it TESS} observations, \cite{2022ApJ...936..187O} have presented a thorough asteroseismic analysis applied to PG 1159-035, the prototype of  GW Vir stars \citep{1979wdvd.coll..377M}. In these works, detailed asteroseismic studies have been carried out, leading to the characterization and determination of the fundamental parameters of these stars. 

\begin{figure*}
\centering
 \includegraphics[clip,width=1.0\linewidth]{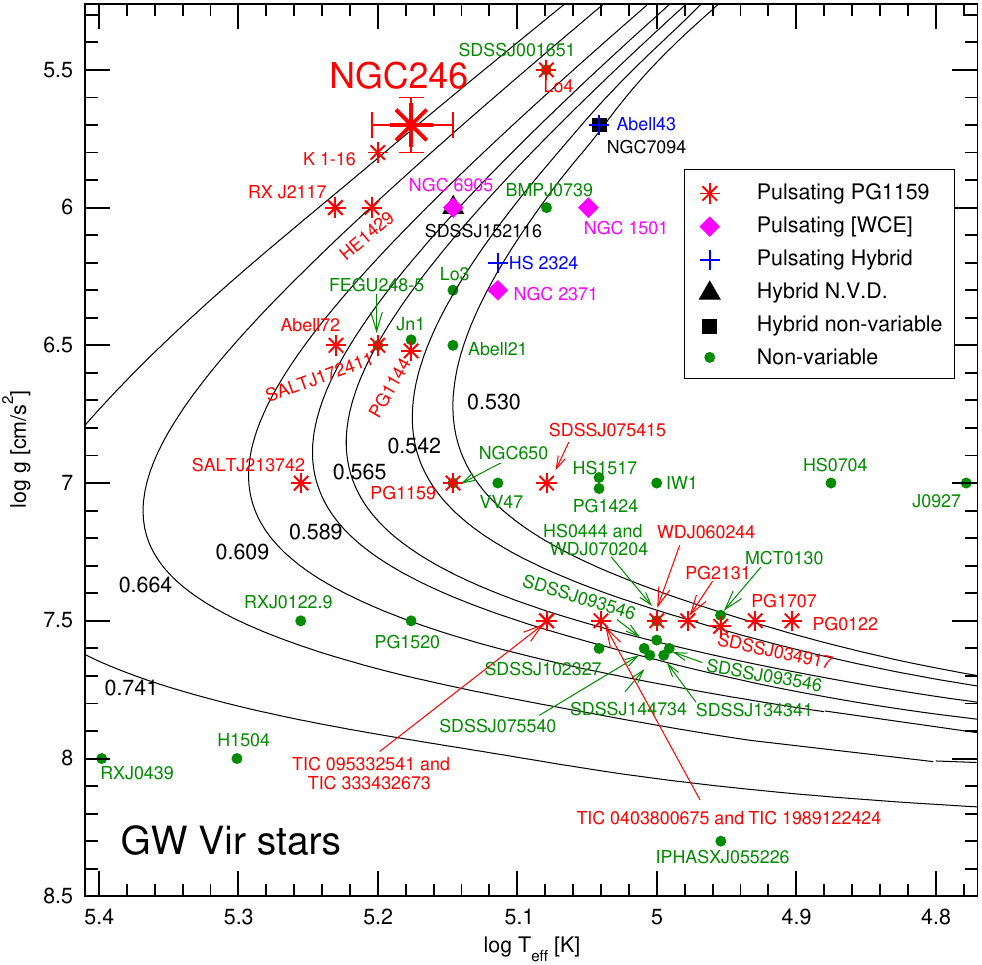}
	\caption{The already known variable and non-variable PG~1159 stars and variable [WCE] stars in the $\log T_{\rm eff}-\log g$ diagram.   Thin solid curves show the post-born again evolutionary tracks from \cite{2006A&A...454..845M} for different stellar masses.  "Hybrid" refers to PG~1159 stars exhibiting H in their atmospheres, where "N.V.D." means no variability data (only applies to SDSS~J152116). The location of the GW Vir star NGC~246 is emphasized with a large red star symbol with error bars.  This star is characterized by $T_{\rm eff}= 150\,000\pm 10\,000$ K and $\log g= 5.7\pm0.10$}.
	\label{fig:logteff-logg}
\end{figure*}

In the present paper, we report new {\it TESS} observations of NGC~246 (TIC\,3905338, according to its {\it TESS} designation), an already-known GW Vir star that is the central star of the planetary nebula NGC~246\footnote{Although NGC~246 is the name of the planetary nebula, throughout this work we use this designation to refer to the central star, for simplicity.}, and carry out a detailed asteroseismic analysis on the basis of the full PG~1159 evolutionary models of \citet{2005A&A...435..631A} and \citet{2006A&A...454..845M}. The NGC~246 planetary nebula has a slightly elliptical shape, and its central star (HIP 3678) is a hierarchical triple stellar system \citep{2014MNRAS.444.3459A}, the only confirmed such system in a planetary nebula. The primary, NGC~246 (a.k.a. HIP 3678 A) is a PG~1159 star with $T_{\rm eff}= 150\,000\ $K and $\log(g)= 5.7$ cgs, according to the non-LTE model atmosphere analysis by \cite{RauchWerner1997}, and $M_{\star}\sim 0.75\ M_{\odot}$ \citep{2006PASP..118..183W,2006A&A...454..845M}. Its orbital companions (HIP 3678 B \& C) are likely low-mass main sequence stars of spectral type K and M \citep{2014MNRAS.444.3459A}. The location of NGC~246  in the $\log T_{\rm eff} - \log g$ diagram is displayed in Fig.  \ref{fig:logteff-logg}, along with other GW Vir stars. A more recent analysis carried out by \citet{Loebling2018,Loebling2020}, employing state-of-the-art non-LTE atmosphere models, considering opacities of 17 chemical species, including iron-group elements, and employing new spectroscopic observations, confirmed the previous results, but with fairly tight error estimates: \Teff = $150\,000\pm10\,000\ $K and $\logg = 5.7\pm0.1$. A comparison with the evolutionary tracks of \cite{2006A&A...454..845M} results in a stellar mass of $0.74^{+0.19}_{-0.23}$\Msol\  \citep{Loebling2018,Loebling2020}. The latest distance measurements resulting from {\sl Gaia} parallax \citep[$\pi= 1.80 \pm 0.08$ mas;][]{2020yCat.1350....0G} places NGC 246 in $d_G= 538^{+20}_{-17}\ $pc \citep{2021AJ....161..147B}.

The reported atmospheric parameters were derived using hydrostatic model atmospheres. This is justified because the star has a weak wind that only affects the formation of metal resonance lines in the ultraviolet wavelength region. Other spectral lines are not affected because they form in the hydrostatic layer of the atmosphere. The \ion{C}{iv} 1548/1551\,\AA\ and \ion{O}{vi} 1032/1038\,\AA\  resonance lines in NGC~246 display prominent P~Cygni profiles, which were analysed with non-LTE wind models by \cite{KoesterkeWerner1998} and \cite{KoesterkeDreizlerRauch1998}. A mass-loss rate of $\log \dot{M}/(M_\odot\,{\rm yr}^{-1}) = -6.9$ was measured.  This value is significantly lower than the mass-loss rates of [WCE] central stars (the putative progenitors of PG1159 stars), for which mass-loss rates have been determined that are on average about an order of magnitude larger. As a consequence, the spectra of [WCE] stars are dominated by emission lines that form in the dense stellar wind. It is worth mentioning that weak stellar winds are not expected to affect the computed pulsation modes as their weight functions are small at the stellar surface during this part of the star’s evolution \citep[see Figs. 5 and 8 of][ respectively]{1985ApJ...295..547K,2006A&A...454..863C}.

In the first literature notes from \citet{1983AJ.....88..439L} and \citet{1987ApJ...323..271G}, this star was listed as a non-variable star. Only a decade later, the observations of \citet{1996AJ....111.2332C} carried out on three nights showed its low-amplitude pulsations. The first two (consecutive) nights revealed significant peaks at 683 and 648 $\mu$Hz in the Fourier spectrum, while the last observation, nine months later, showed a different peak at 540 $\mu$Hz. A more extensive study  conducted by \citet{2003BaltA..12..125G} (three short runs carried out in the years 2000-2001) showed a complex and highly variable short-scale photometric behavior. In the 2000 observing run, the first night ($2.2\ $h long) showed a peak around $550\ \mu$Hz, while the second ($3\ $h, three days later) showed a lower amplitude peak in the same region and, also, a stronger peak at $685\ \mu$Hz. Observations from over a year later (more than $4\ $h long) showed a peak at $\sim 450\  \mu$Hz and additional possible peaks located around $400$ and $500\ \mu$Hz. The last run was carried out two months later and, because of its poorer quality, it only showed low-amplitude peaks at the same region as in the previous run. Moreover, the authors claimed possible features in the Fourier spectra that may be related to interactions with a close companion (for instance, alleged harmonics of the $450\ \mu$Hz peak). All these observations, carried out over more than two decades, only proved that in order to reveal the details of the complex nature of the pulsations of the central star in NGC~246, extensive monitoring on a longer time base was very necessary. With a lack of a dedicated Whole Earth Telescope (WET) campaign \citep{1990ApJ...361..309N}, as already suggested by \citet{2003BaltA..12..125G}, the \textit{TESS} observations finally give us the exciting opportunity to study the complex behavior of the GW Vir star in NGC~246 in detail.
A preliminary analysis of this star has been previously presented by \cite{2020A&A...635A.128A}.

The present work is the sixth in the series of papers dedicated to studying pulsating H-deficient WDs and pre-WDs observed with {\it TESS}. The first paper was focused on the analysis of six already known GW Vir stars \citep{2021A&A...645A.117C}, and the second one, on the discovery of two new stars of this type \citep{2021A&A...655A..27U}. In the third part, \cite{2022A&A...659A..30C} analyzed the DBV star\footnote {DBVs are He-rich atmosphere pulsating WD stars with effective temperatures and surface gravities in the ranges $22\,400 \lesssim T_{\rm eff} \lesssim 32\,000$ K and $7.5 \lesssim \log g \lesssim 8.3$, respectively \citep{2019A&ARv..27....7C}.}  GD~358, while in the fourth \citep{2022MNRAS.513.2285U}, the discovery of two additional GW Vir stars was reported. In the fifth paper \citep{2022A&A...668A.161C}, the authors presented the discovery of two new DBV pulsators and the analysis of three already-known pulsating WD stars of this type.

This paper is organized as follows. In Sect.~\ref{photometry}, we provide details of the observations and reduction of the photometric {\it TESS} data and also, on the frequency analysis. In Sect.~\ref{astero-analyses}, we start by giving a brief summary of the evolutionary models and numerical tools employed for our theoretical analyses. Next, we focus on the asteroseismic modeling of NGC~246, including the search for a possible mean period spacing within the period spectra (Sect.~\ref{period-spacing}), the derivation of the stellar mass using the inferred period spacing (Sect.~\ref{mass-period-spacing}), and the search for a period-to-period fit with the aim of finding an asteroseismic model that best represents the pulsations of the target star (Sect.~\ref{period-to-period-fits}). Finally, in Sect.~\ref{conclusions}  we summarize our main findings.

\section{Observations and data reduction}
\label{photometry}

NGC~246 was observed by {\it TESS} in Sector 3 between September 20 and October 18, 2018, and in Sector 30 between September 22 and October 21, 2020. In Sector 3, this star was observed during 20.27 days with the short cadence (SC, $120\ $s) mode, with a 1.19-day gap between the orbits, which is shorter than the 27-day nominal duration of one-sector observations, while in Sector 30 it was observed for 25.92 days with a 1.07-day gap in both SC and ultra-short cadence mode (USC, $20\ $s). There is a 709.67-day data gap between the two sectors.

We first investigated the target-pixel files (TPFs) from {\it TESS} to assess the potential contamination from the nearby objects within the {\it TESS} aperture by using different aperture shapes, sizes, and systematics corrections. We downloaded TPFs for Sectors 3 and 30 in \textsc{Lightkurve} \citep{2018ascl.soft12013L} for further processing. Then, we created custom apertures and used them to extract raw light curves from the TPFs. TPFs are already processed by the Science Processing Operations Center (SPOC) pipeline \citep{Jenkins2016} that removes backgrounds, but it is still possible to remove common trends in TPF pixels. We found that Pixel Level Decorrelation (PLD) \citep{2015ApJ...805..132D} within \textsc{Lightkurve} gave the best results compared to the Pre-search Data Conditioning Simple Aperture Photometry (PDCSAP) flux. PLD is a special case of a Regression Corrector that uses linear regression to remove systematic noise from light curves based on a design matrix created using only background pixels surrounding the target aperture. A noise model created using this information from nearby pixels is then subtracted from the uncorrected light curve. %.

\begin{figure}
	\includegraphics[width=\columnwidth]{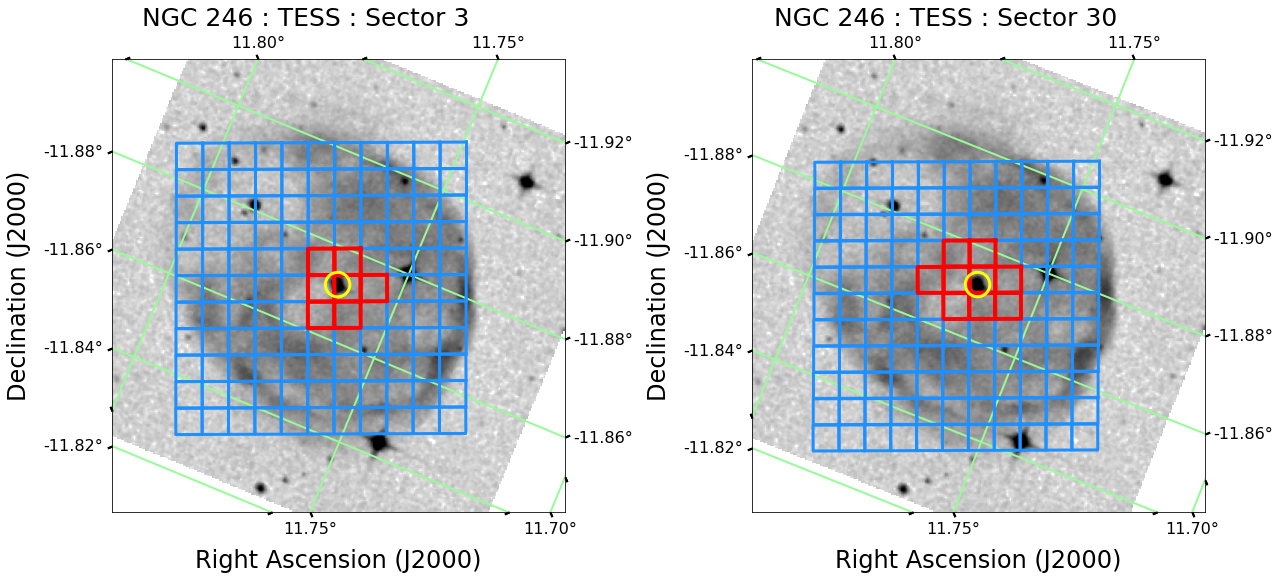}
	\caption{TESS Sectors 3 and 30 pipeline aperture overlaid on DSS2 Red sky image. The target is marked with yellow circles. Blue squares show the TPF overlay and red squares show the TPF aperture overlay for each Sector. Figure produced using \texttt{mkpy3} by Kenneth Mighell (\url{https://github.com/KenMighell/mkpy3}).} 
	\label{fig:tessoverplotted}
\end{figure}

In Fig.~\ref{fig:tessoverplotted}, we show the pipeline apertures used to extract photometry in Sectors 3 and 30, overlaid on the DSS2 Red image. It is evident that nebular contamination is small but inevitable, and it varies between the two sectors because of the different sizes of the apertures. The on-sky orientation of the field is almost the same (there is only a small, $\sim 1\ $pix shift), therefore the use of an aperture of the same shape (but slightly shifted) would be justified and perhaps more appropriate in this case.

\begin{figure}
	\includegraphics[width=\columnwidth]{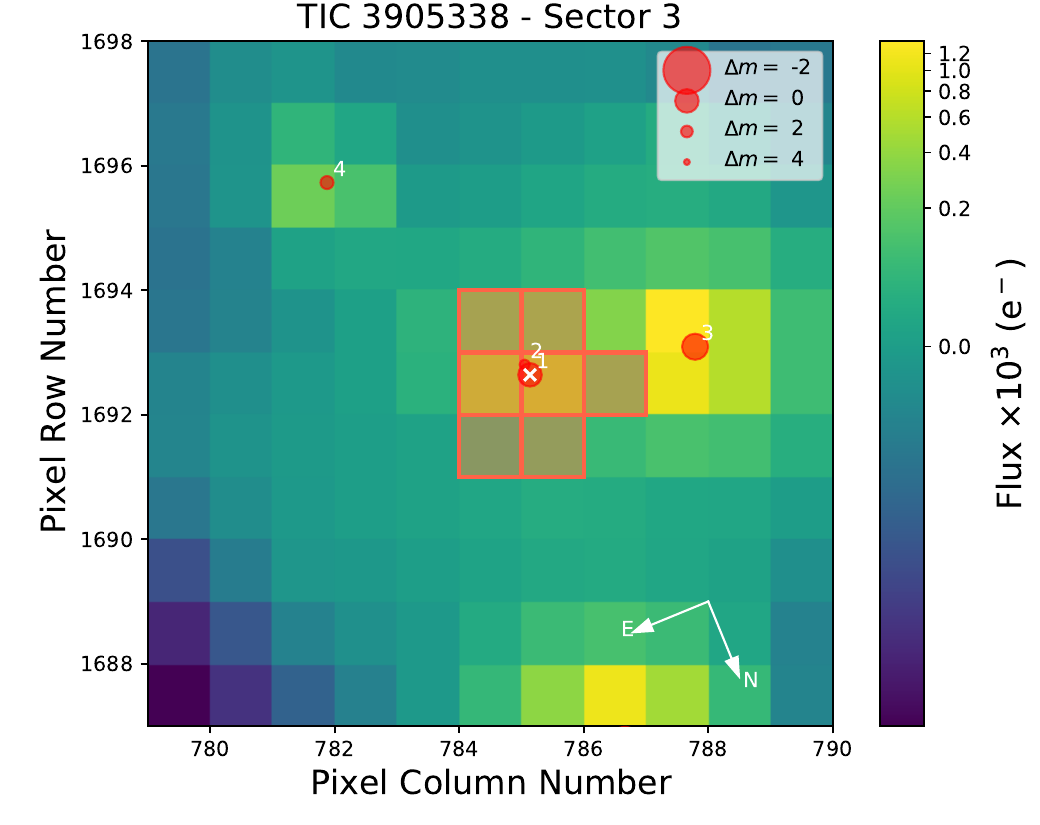}
    \caption{TESS Target Pixel File (TPF) with the position of NGC\,246 marked with a white ``X''. The other sources in the field from Gaia DR2 are marked with red circles scaled with magnitudes. The aperture mask used by the PDC pipeline is shown with red squares.}
    \label{fig:tesscontam}
\end{figure}

%. 
Figure~\ref{fig:tesscontam} shows the target ($T_{mag}=12.31$) with other stars from the
{\sl TESS} input catalog \citep{Stassun2019}. There are two main sources of contamination: the $T_{mag}=13.70$ star close to the target position which is the wide binary member of the NGC 246 triple system (HIP 3678 B), and the $T_{mag}=11.14$ star, TYC 5272-1854-1, to the right. The TESS header contains two parameters for crowding correction: \texttt{CROWDSAP} - the crowding metric, defined as the ratio of the target flux to the total flux in the optimal aperture, and \texttt{FLFRCSAP} - the flux fraction, which reflects what fraction of the Pixel Response Function of the target is outside the optimal aperture (missing flux). In the case of NGC~246, the \texttt{CROWDSAP} value is $0.72$ for Sector 3 and $0.64$ for Sector 30, while \texttt{FLFRCSAP} is $0.72$ and $0.78$, respectively. It means that using solely the \texttt{CROWDSAP}, only 72 and 64\% of the total flux originally measured in the aperture originates from the target in Sectors 3 and 30, respectively. The possible influence on the variability of NGC 246 by the 13.70 mag star is hard to assess with only TESS data at hand, therefore we checked whether the 11.14 mag star shows any variability in TESS data. We varied aperture sizes and shapes for that target and concluded that there was no variability from that star that could influence our analysis of NGC~246. With the ground-based data from \citet{2006A&A...454..527G} reduced anew, using only a short run where it was possible to resolve the stars, we were not able to verify the lack of variability in the 13.70 mag star on time scales relevant to our analysis. Since this star, HIP 3678 B, is an early-to-mid K dwarf star \citep{2014MNRAS.444.3459A}, it is not expected to exhibit variability that we could confuse with pulsations of the white dwarf primary. Our new analysis of this data set only confirms the findings reported by the authors that the Fourier spectrum was variable on a nightly basis.

For the analysis presented in the remainder of the paper, we extracted times in Barycentric corrected Julian days (''BJD-245700'') and fluxes already corrected for contamination by the pipeline (''PDCSAP FLUX'') from the FITS files using \textsc{Lightkurve}.
The fluxes were normalized and transformed to amplitudes in parts-per-thousand (ppt) units.
To remove the outliers that appear above 5 times the median of intensities, the data were finally sigma-clipped based on 5$\sigma$.

\subsection{Frequency analysis}
\label{sect:freq_analysis}

%S30 = 25.9264764 days - 95873 points

\begin{figure} 
\includegraphics[clip,width=1.\columnwidth]{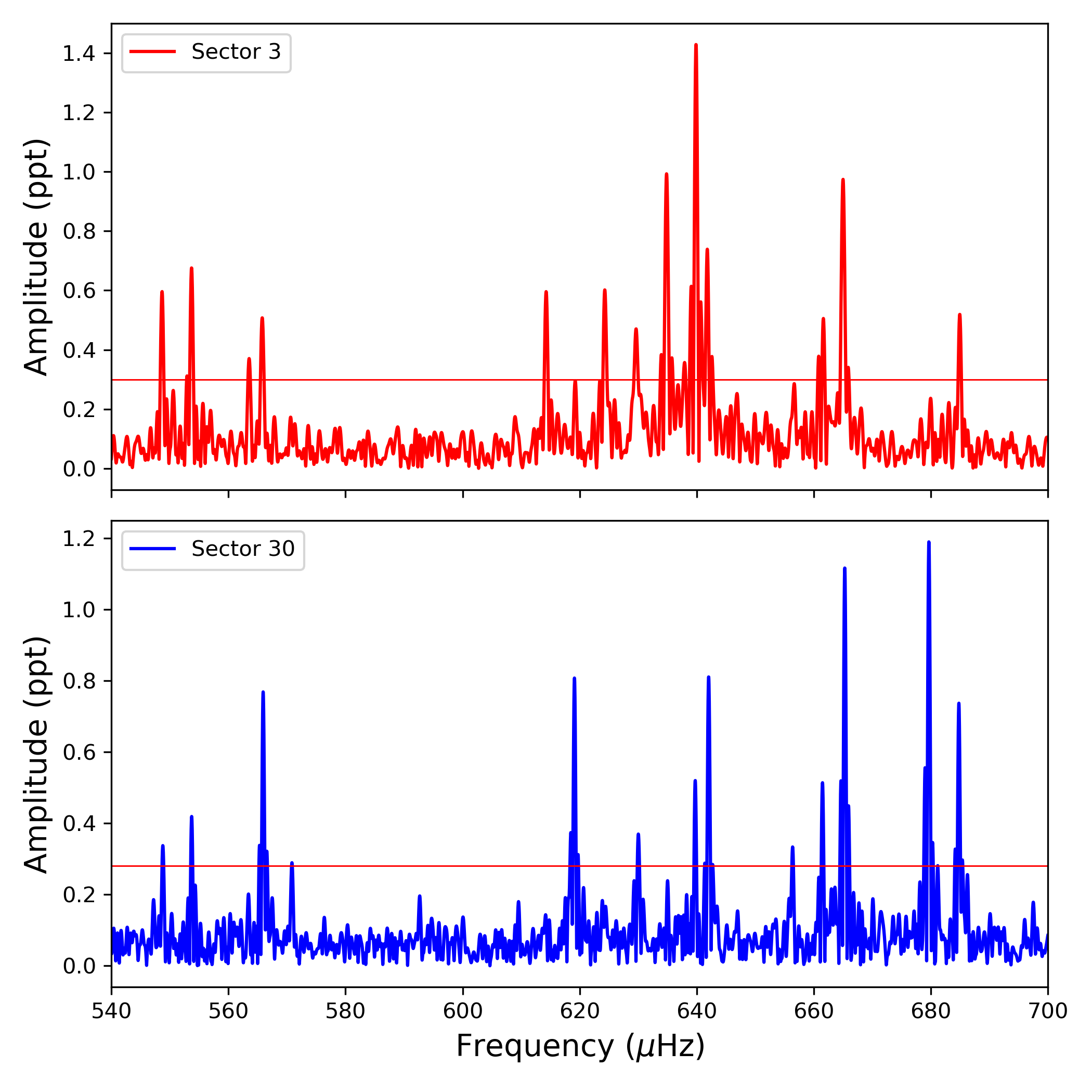}
\caption{Fourier transform of Sector 3 ({\it upper panel}) and Sector 30 ({\it lower panel}) of NGC~246. The horizontal red line indicates the 0.1\% FAP level.%.
}
\label{fig:FT} 
\end{figure}

In order to analyze the periodicities in the data and determine the frequency of each pulsation mode, along with their amplitude and phase, Fourier transforms (FT) of the light curves were computed.
We choose a 0.1 percent false alarm probability (FAP) detection threshold, which means that if the amplitude exceeds this threshold, there is a 0.1 percent chance that it is merely the result of noise fluctuations.
Following the procedure outlined in \citet{kepler1993}, we determined the $0.1\%$ FAP threshold.
We performed a nonlinear least square (NLLS) fit in the form of $A_i \sin(\omega_i\ t + \phi_i)$, with $\omega=2\pi/P$, where $P$ is the period.
%. 
We identified the frequency (period), phase, and amplitude values for each periodicity in this way.
We prewhitened the light curves using the NLLS fit parameters until, excepting unresolved peaks, there was no longer any significant signal in the FT of both datasets above the significance level of $0.1\%$ FAP.
%.
All prewhitened frequencies for NGC\,246 are given in Table \ref{F_list}, where we present frequencies (periods) and amplitudes, along with their corresponding errors and the S/N ratio.
The average noise level of Sector 3 corresponds to 0.06 ppt and the significance level of $0.1\%$ FAP, to 0.29 ppt. We extracted 15 frequencies above 0.29 ppt from the light curve, as shown in the top panel of Fig. \ref{fig:FT}. All the pulsational frequencies are located between 548 $\mu$Hz (1822 s) and 684 $\mu$Hz (1459 s). 
%.
For Sector 30, the average noise level corresponds to 0.055 ppt. In total, we detected 12 pulsational frequencies above the detection threshold of 0.26 ppt. This case is shown in the bottom panel of 
Fig. \ref{fig:FT}. %. 
In Table \ref{F_list}, we reported the final best-fit values for the frequencies (periods) and amplitudes of the individual sinusoids in our model.
%.

\subsection{Possible rotational multiplets?} 
\label{rotational}

As we will see, identification of the value of $\ell$ is a critical component of an asteroseismic analysis.  One property of nonradial pulsations in compact stars -- the effects of rotation -- often provides critical data to identify the value of $\ell$ (and $m$). For a nonrotating star, the pulsation frequencies for modes with the same value of $k$ are the same for all values of $\ell$ and $m$.  Rotation, however, can lift that degeneracy, splitting modes with different sets of ($\ell$, $m$) by amounts proportional to the rotation frequency $\Omega$.  As long as the rotation frequency is significantly smaller than the pulsation frequency (i.e. to first order in $\Omega/\sigma_k$), we have:
\begin{equation}
\label{eq:rotsplit}
    \sigma_{k, \ell, m} = \sigma_{k} + m \Omega (1 - C_{k,\ell})
\end{equation}
where $\sigma_k$ is the oscillation (angular) frequency in the nonrotating case, and $C_{k,\ell}$ is a function (integrated through the star) of the stellar structure and the eigenfunctions for the mode \citep[for details see e.g.][]{Aerts2010}.  

Thus, rotation can produce a triplet of $\ell=1$ modes, with a central peak ($m=0$) flanked by peaks separated on either side by $\Omega (1-C_{k,\ell})$.  Higher $\ell$ modes can be split into multiplets with $2 \ell + 1 $ components, split equally by the same factor (though the value of $C_{k,\ell}$ depends on $\ell$).

For $g$ modes in compact stars, where the horizontal displacement is much larger than the radial displacement, the value of $C_{k, \ell}$ is closely approximated by \citep{1991ApJ...378..326W}:
\begin{equation}
\label{eq:csimp}
    C_{k, \ell} = \frac{1}{\ell (\ell+1)}.
\end{equation}
So for $\ell = 1$ modes, the (angular) frequency spacing between peaks in the FT is $\Omega/2$, while for $\ell = 2$ modes we would (ideally) see a quintuplet of modes, split by $\Omega \times 5/6$.  The rotational splitting of all modes (all values of $k$) of a given $\ell$ should be identical unless there is latitudinal differential rotation.  If both $\ell=1$ and $\ell=2$ modes are excited in the star, the values of their rotational splitting should be related by exactly $\frac{\Delta f_{\ell=1}}{\Delta f_{\ell=2}}=0.6$. 

If the rotational period is not longer than the duration of the observation, this 2$\ell + 1$ configuration can be resolved using high-precision photometry. Rotational multiplets have been detected for all types of pulsating WDs, including GW Vir, DBV, DAV\footnote{DAVs are H-rich atmosphere pulsating WD stars with effective temperatures and surface gravities in the ranges $10\,400 \lesssim T_{\rm eff} \lesssim 12\,400\ $K and $7.5 \lesssim \log g \lesssim 9.1$, respectively \citep{2019A&ARv..27....7C}.}, and ELMV\footnote{ELMVs are H-rich atmosphere pulsating Extremely Low-Mass (ELM) WD stars with effective temperatures and surface gravities in the ranges $7\,800 \lesssim T_{\rm eff} \lesssim 10\,000$ K and $6.0 \lesssim \log g \lesssim 6.8$, respectively \citep{2019A&ARv..27....7C}.} stars, and the derived rotation periods range from 1 hour to a few days \citep{2017ApJS..232...23H,2019A&ARv..27....7C,2021ApJ...922..220L,2022A&A...668A.161C,2022MNRAS.513.2285U,2023A&A...673A.135C}.

For a rotation period of 1 or 2 days, the $\ell=1$ splitting corresponds to 5.79 or 2.89$\,\mu$Hz, respectively.  The $\ell=2$ splitting is larger (see Equation \ref{eq:csimp}) by a factor of 5/3. 
In most of the cases cited above, the frequency splitting between peaks in a given ($k, \ell$) multiplet is significantly smaller than the spacing between modes of different $k$.  But for pulsating PN central stars, which have periods that are significantly longer, the separation in frequency between modes of different $k$ is smaller.
For the long periods as those detected in NGC 246, the period spacing between consecutive $m=0$ modes results in a small frequency spacing which can be confused with the rotational splitting, so care must be exercised to separate the two effects \citep[similar issues arose in the study of the central star of NGC~1501 by ][]{1996AJ....112.2699B}.

With this theoretical background, we attempted a search for frequency splittings within the detected frequencies.  First, we searched for pairs of peaks with a similar frequency separation. Frequencies f$_{10,S03}$ and f$_{11,S03}$ differ by $641.829-639.872=1.957\,\mu$Hz, while frequencies f$_{14,S03}$ and f$_{15,S03}$ by $664.997-661.711 = 3.286\,\mu$Hz. The relation between those separations is almost exactly $0.6$, therefore we might suspect that f$_{10,S03}$ and f$_{11,S03}$ could be $\ell=1$ modes, whereas f$_{14,S03}$ and f$_{15,S03}$ could be $\ell=2$ modes, both cases with missing components. Another pair of $\ell=1$ modes could be f$_{3,S03}$ and f$_{4,S03}$: $565.800-563.500=2.300\,\mu$Hz. 
On the other hand, the frequency splitting might be larger, as we found four sequences split by approximately $5\,\mu$Hz (it could even be a sequence of five modes, f$_{6}-\mathrm{f}_{10}$). Those are: \\
f$_{1,S03}$ and f$_{2,S03}$: $553.725-548.646=5.079\,\mu$Hz, \\
f$_{6,S03}$ and f$_{7,S03}$: $624.294-619.230=5.064\,\mu$Hz,  \\
f$_{9,S03}$ and f$_{10,S03}$: $639.872-634.850=5.022\,\mu$Hz, and \\
f$_{13,S30}$ and f$_{14,S30}$: $661.506-656.412=5.094\,\mu$Hz.\\
If those modes were $\ell=1$ we could expect the $\ell=2$ modes split by about $8.4\,\mu$Hz, otherwise we could expect the $\ell=1$ modes split by about $3\,\mu$Hz. We did not find any clear sequence with those values other than already mentioned, therefore we explored a possibility of overlapping mode sequences with missing components. For example, if we consider the difference between f$_{7,S03}$ and f$_{11,S03}$ of $641.829-624.294=17.535\,\mu$Hz, it could be a triplet of $\ell=2$ modes with a missing component in between split by $8.767\,\mu$Hz (and missing side components), overlapping in the same region as previously mentioned $\ell=1$ modes split by about $5\,\mu$Hz. A multitude of options prevents us from a definitive determination of the rotation period in NGC 246 and mode identification based on the observed frequency splitting.

\begin{table*}
\setlength{\tabcolsep}{4.8pt}
\renewcommand{\arraystretch}{1.1}
\centering
\caption{Frequency solution from the light curves of Sectors 3 and 30 of NGC~246 including frequencies, periods, 
amplitudes (and their uncertainties), and S/N. }

\begin{tabular}{cccccccccr}
\hline
\noalign{\smallskip}
ID & Frequency (S03) & Period (S03) & Amplitude (S03) & S/N & Frequency (S30) & Period (S30) & Amplitude (S30) & S/N  \\
   & $\mu$Hz   &  [sec] &  [ppt]  &  &    $\mu$Hz   &  [sec] &  [ppt]    &       \\
\hline
\noalign{\smallskip}

f$_{\rm 1}$ &  548.646 (27) &   1822.67 (9) &   0.59 (7) &   9.2  & 548.774 	(39) &	1822.243	(13) &	0.34	(4) &	5.0      \\
f$_{\rm 2}$ &  553.725 (22) &   1805.95 (7) &   0.67 (7) &  11.2  & 553.701 	(26) &	1806.026	(08) &	0.42	(4) &	7.6   \\
f$_{\rm 3}$ &  563.500 (46) &   1774.62 (14) &   0.37 (7) &   5.4 \\
f$_{\rm 4}$ &  565.800 (32) &   1767.41 (10) &   0.52 (7) &   7.6   & 565.929 	(13) &	1767.005	(04) &	0.77	(4) &	15.1  \\
f$_{\rm 5}$ &  614.268 (25) &   1627.96 (7) &   0.61 (7) &   9.8 \\
f$_{\rm 6}$ &  619.23 (5) &   1614.90 (14) &   0.30 (7) &   4.6 &  619.126 	(13) &	1615.180	(03) &	0.81	(4) &	15.1    &\\
f$_{\rm 7}$ &  624.294 (25) &   1601.81 (6) &   0.61 (7) &   9.9 \\
f$_{\rm 8}$ &  629.727 (32) &   1587.99 (8) &   0.48 (7) &   7.7 & 630.058 	(29) &	1587.154	(07) &	0.33	(4) &	6.8   \\
f$_{\rm 9}$ &  634.850 (15) &   1575.175 (37) &   1.00 (7) &  16.7  &   \\
f$_{\rm 10}$ &  639.872 (10) &   1562.813 (25) &   1.44 (7) &  24.4  &   639.793 	(19) &	1563.004	(04) &	0.51	(4) &	10.2   \\
f$_{\rm 11}$ &  641.829 (18) &   1558.048 (45) &   0.75 (7) &  13.3  &   642.042 	(13) &	1557.529	(03) &	0.81	(4) &	14.4  \\
f$_{\rm 12}$ &  646.78 (5) &   1546.13 (12) &   0.32 (7) &   4.9 \\
f$_{\rm 13}$ &   &   &  &  &   656.412 	(43) &	1523.431	(10) &	0.33	(4) &	4.6    \\  
f$_{\rm 14}$ &  661.711 (25) &   1511.23 (6) &   0.51 (7) &   9.9 &  661.506 	(20) &	1511.700	(04) &	0.52	(4) &	9.8        \\
f$_{\rm 15}$ &  664.997 (15) &   1503.766 (34) &   0.98 (7) &  16.4  &  665.305 	(09) &	1503.068	(02) &	1.12	(4) &	21.7       \\
f$_{\rm 16}$ &   &   &   &   &  679.693 	(08) &	1471.251	(01) &	1.19	(4) &	23.1  \\
f$_{\rm 17}$ &  684.955 (27) &   1459.95 (6) &   0.52 (7) &   8.9  &  684.795 	(15) &	1460.289	(03) &	0.74	(4) &	13.3       \\
\hline
\noalign{\smallskip}
\label{F_list}
\end{tabular}
\end{table*}

\section{Asteroseismic analysis}
\label{astero-analyses}

The asteroseismic analysis performed in this work is based on a set of stellar models that take the complete evolution of the PG~1159 progenitor stars into account \citep{2005A&A...435..631A,2006A&A...454..845M,2007A&A...470..675M,2007MNRAS.380..763M}. Post-AGB evolutionary sequences, computed with the {\tt LPCODE} evolutionary code \citep{2005A&A...435..631A}, were followed through the very late thermal pulse (VLTP) episode and the resulting born-again episode that leads to the H-deficient, He-, C-, and O-rich composition expected for PG~1159 stars. The resulting stellar remnants have masses of $0.515$, $0.530$, $0.542$, $0.565$, $0.589$, $0.609$, $0.664$, $0.741$, and $0.870\ M_{\sun}$. In Fig~\ref{fig:logteff-logg}, the evolutionary tracks corresponding to the PG~1159 models employed in this work are shown in the $\log(T_{\rm eff})-\log(g)$ plane. Details about the input physics and evolutionary calculations carried out to obtain the PG~1159 evolutionary sequences employed in this work can be found in  \citet{2005A&A...435..631A,2006A&A...454..845M,2007A&A...470..675M,2007MNRAS.380..763M}. 
We computed $\ell= 1, 2$ $g$-mode adiabatic pulsation periods in the range $80$ - $6000\ $s with the adiabatic version of the pulsation code {\tt LP-PUL} \citep{2006A&A...454..863C}. We analyzed roughly $4000$ PG~1159 models covering a wide range of effective temperatures ($4.8 \lesssim \log T_{\rm eff}  \lesssim 5.4$), luminosities ($0 \lesssim \log (L_{\star}/L_{\odot}) \lesssim 4.2$), and stellar masses ($0.515 \leq M_{\star}/M_{\odot} \leq 0.870$).

The nonradial $g$ modes that cause brightness variations in WDs and pre-WDs can be excited in a sequence of consecutive radial orders, $k$, for a given value of the harmonic degree, $\ell$. In the asymptotic limit of high radial order, the periods of $g$ modes with consecutive radial orders are approximately evenly separated, being the period spacing dependent on $\ell$, according to the expression  \citep{1990ApJS...72..335T}:

\begin{equation}
\label{eq:spacing}
\Delta\Pi^{\rm a}_{\ell}= \Pi_{k+1,\ell}-\Pi_{k,\ell}= \frac{\Pi_0}{\sqrt{\ell(\ell+1)}},
\end{equation}

\noindent where $\Pi_0$ is a constant value given by:

\begin{equation}
\Pi_0=\frac{2\pi^2}{\left[\int^{r_2}_{r_1}{\frac{N}{r}dr}\right]}
\end{equation}

\noindent and $N$  is  the  Brunt-V\"ais\"al\"a  frequency. In the case of chemically homogeneous stellar models, this formula for the asymptotic period spacing ($\Delta\Pi^{\rm a}_{\ell}$) represents a precise description of the separation of consecutive $g$-mode periods for large radial orders. A very relevant property of the $g$-mode period spacing of pulsating PG~1159 stars is that it primarily depends on the stellar mass, and only weakly on the luminosity and the He-rich envelope mass  \citep{1986PhDT.........2K,1987fbs..conf..297K,1988IAUS..123..329K,1990ASPC...11..494K,1994ApJ...427..415K,2006A&A...454..863C}. This property, in principle, can be used to derive the mass of a star by measuring the period spacing.
For chemically stratified stars, such as PG~1159 stars, the $g$-mode period spacings considerably depart from uniformity due to the mechanical resonance called “mode trapping” \citep{1994ApJ...427..415K}. Therefore, at first glance,  it would seem that the observed period separation of a chemically stratified star could not be used to infer its stellar mass. Fortunately, the {\it average} value of the period spacings in chemically layered PG~1159 stars still preserves the property of being a function almost exclusively of the stellar mass \citep{1990ApJS...72..335T}. Thus, one can still derive the stellar mass of chemically stratified stars from the comparison between the observed mean spacing of periods, $\Delta\Pi$, and the average of the period spacings, $\overline{\Delta  \Pi}$, calculated for sets of PG~1159-star models with different masses and effective temperatures.

Another approach to deriving the stellar mass, which, at the same time, can give information about the internal structure of the pulsating star, is to search for a pulsation model that best matches the \emph{individual} pulsation periods of the star under study. The goodness of  the  match  between the   theoretical pulsation  periods ($\Pi_k^{\rm  T}$) and  the observed   individual  periods ($\Pi_i^{\rm  O}$) is measured by means  of a merit function defined as 

\begin{equation}
\label{chi}
\chi^2(M_{\star},  T_{\rm   eff})=   \frac{1}{m} \sum_{i=1}^{m}   \min\left[\left(\Pi_i^{\rm   O}-   \Pi_k^{\rm  T}\right)^2\right], 
\end{equation}

\noindent where $m$ is the number of observed periods. The PG 1159 model that shows the lowest value of $\chi^2$, if exists, is adopted as the ``best-fit model''. We assess the function $\chi^2=\chi^2(M_{\star}, T_{\rm eff})$ for stellar masses of $0.515$, $0.530$, $0.542$, $0.565$, $0.589$, $0.609$, $0.664$, $0.741$, and $0.870\ M_{\sun}$.  For the effective temperature, we employ a much finer grid ($\Delta T_{\rm eff}= 10-30$ K) which is given by the time step adopted by our evolutionary calculations.

The mentioned methods to estimate the stellar mass and the internal structure of NGC\,246 are the same ones we employed in our previous works at the La Plata Stellar Evolution and Pulsation Research Group\footnote{\url{http://evolgroup.fcaglp.unlp.edu.ar/publications.html}}, for instance, \cite{2007A&A...461.1095C,2007A&A...475..619C,2008A&A...478..869C,2009A&A...499..257C,2014MNRAS.442.2278K,2016A&A...589A..40C,2021A&A...645A.117C,2021A&A...655A..27U}, where further details can be found.

\subsection{Period spacing}
\label{period-spacing}

\begin{figure}  
\centering   \includegraphics[clip,width=270pt]{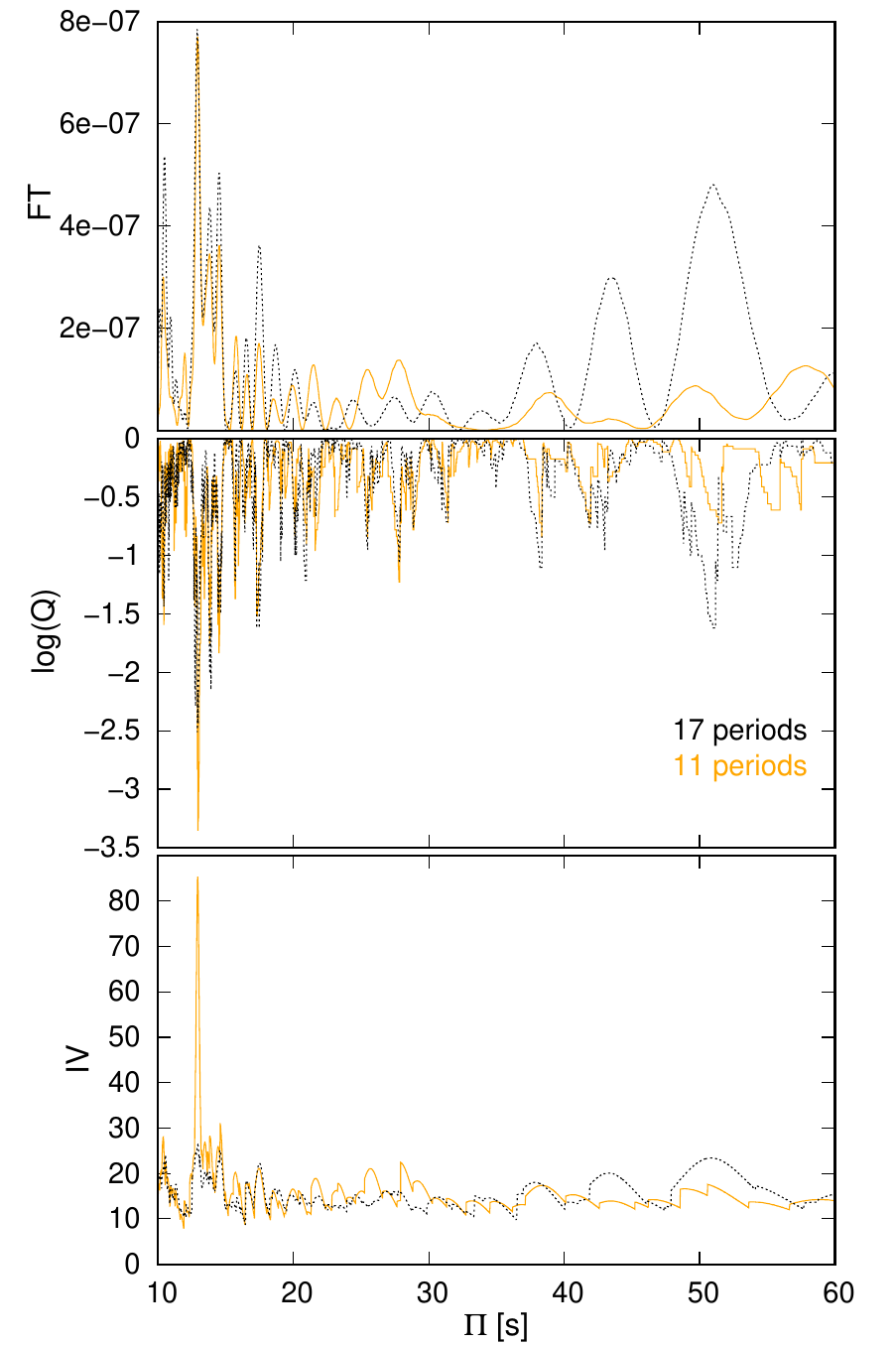} 
\caption{ F-T (\textit{upper panel}), K-S (\textit{middle panel}) and I-V (\textit{bottom panel}) significance  tests applied to the period  spectrum of NGC~246 to search for a constant period spacing. Dotted black lines represent the results for the set of 17 periods from Table~\ref{tab:avrg-sectors}, while solid orange lines, the case for a subset of 11 periods (see text for details).}
\label{fig:tests1}
\end{figure}

As a first step, we aim to estimate a mean period separation underlying the observed periods (if it exists). Considering the set of periods from Table~\ref{F_list}, we searched for a constant period spacing within the data of NGC\,246 employing the Kolmogorov-Smirnov \citep[KS; see][]{1988IAUS..123..329K}, the inverse variance \citep[I-V; see][]{1994MNRAS.270..222O}, and the Fourier transform \citep[F-T; see][]{1997MNRAS.286..303H} significance tests.
Given that, in principle, there are two different sets of periods coming from two sectors, we followed this procedure for each set of periods separately. We additionally considered a combination of both sets, taking the average value for those cases where two periods are identified as the same mode. We show the resulting list of 17 periods in the first column of Table~\ref{tab:avrg-sectors}. Since we found similar solutions for the three cases, we only show the results for the case of the 17 periods from the combined sectors (Table~\ref{tab:avrg-sectors}) in Fig.~\ref{fig:tests1}, indicated with dotted black lines. The figure shows a clear signature of a potential mean period spacing around $12.9\ $s, while other possible values are around $10.6$, $14.5$, and $51\ $s. Next, we repeated this analysis but employing other sets of periods. % 
In our trials, we found that, for instance, when discarding the periods at around $1503$, $1511$, $1557$, $1562$, $1767$, and $1774\ $s, the three statistical tests applied to the resulting set of 11 periods show more robust evidence of a mean period spacing at $\sim 12.9\ $s, as depicted with solid orange lines by Fig.~\ref{fig:tests1}, while the other possible values show much lower (or no) statistical significance, hence, we disregard them. Also, the addition or removal of some of these excluded periods gives approximately the same results, in some cases even improving the statistical significance of the tests. In this way, we adopt the value of $\sim 12.9\ $s as a guess value for the mean period spacing.

We cannot know in advance if this possible mean period spacing is associated with $\ell= 1$ or $2$ modes. We might assume it is a period spacing of $\ell= 2$ modes for its short value, but then a period spacing of $22.3$ ($= 12.9 \times \sqrt 3$, according to Eq.~(\ref{eq:spacing})) corresponding to $\ell= 1$ modes would be expected, which is absent. However, there is no reason to discard this $12.9\ $s period spacing for a sequence of $\ell= 2$ modes. In the following section, we estimate the stellar mass of NGC\,246 by considering the possibility that $\Delta \Pi= 12.9\ $s is associated with $\ell= 1$ or $\ell= 2$ modes.

\begin{figure}  
\centering   \includegraphics[clip,width=270pt]{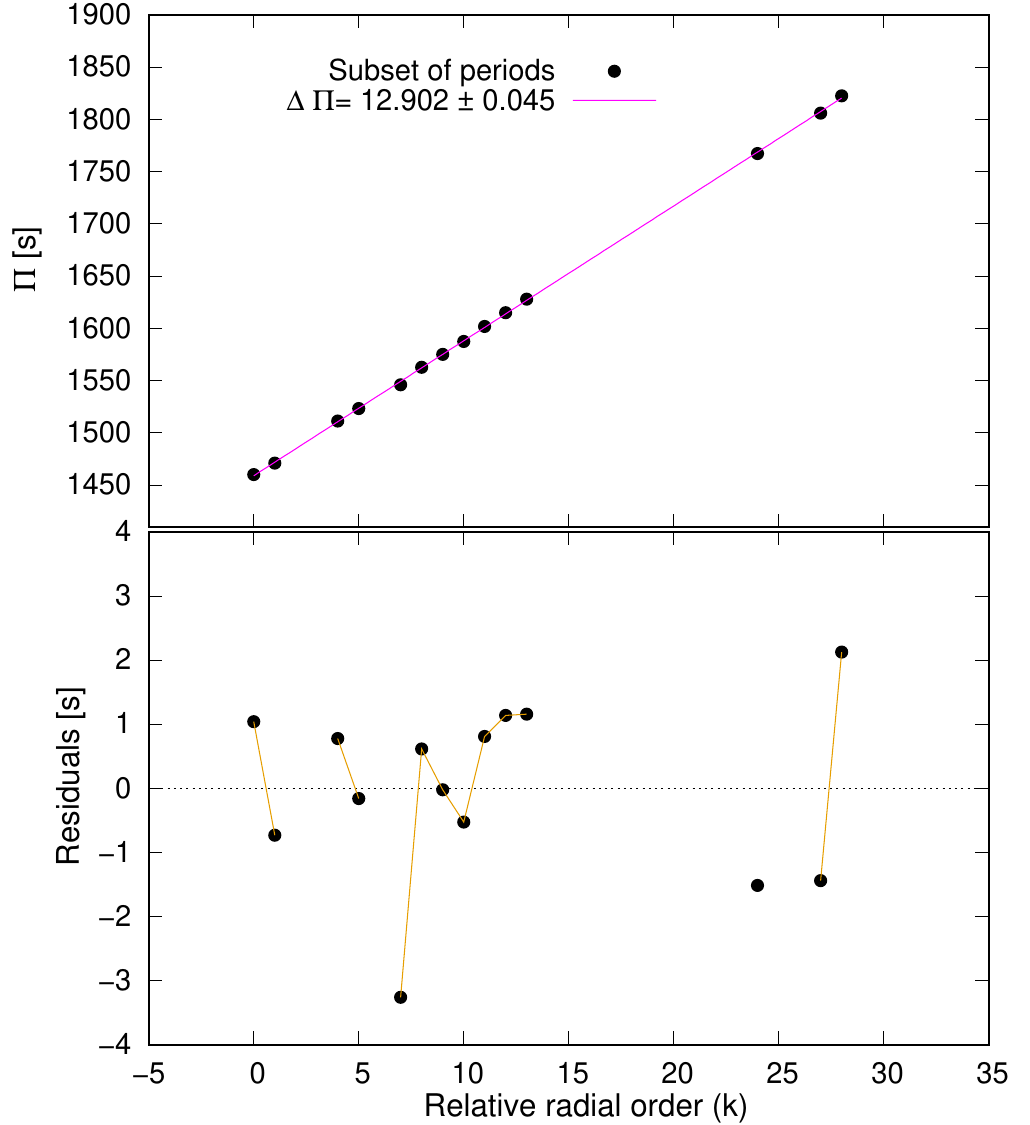} 
\caption{\textit{Upper panel:} Linear least-squares fit applied to a subset of 14 periods of NGC~246 (see text for details). The resulting period spacing is $\Delta \Pi= 12.902 \pm 0.045\ $s. \textit{Lower panel:} Residuals of the period distribution relative to the dipole mean period spacing. A thin orange line connects modes with consecutive radial order.}
\label{fig:linear-fit}
\end{figure}

To derive a refined value of the period spacing first we carried out a linear least-squares fit, using the reduced set of 11 periods. %. 
 We obtained $\Delta \Pi$ = $12.934 \pm 0.054\ $s. The average value of the residuals ($\delta \Pi$) resulting from the difference between the observed periods and those calculated from the mean period spacing is $1.093\ $s. We repeated the linear least-squares fit, but this time, including three of the previously discarded periods ($1511$, $1562$, and $1767\ $s; that is, we used the 14 periods marked with an asterisk in Table~\ref{tab:avrg-sectors}) that are compatible with the previous linear fit, and we obtained $\Delta \Pi$ = $12.902 \pm 0.045\ $s. This value of the period spacing is slightly shorter than the one derived previously, and the uncertainty is slightly smaller. Additionally, the average of the residuals is smaller, $0.986\ $s, so the fitted periods match the observed periods better than before, and this is why we adopt $\Delta \Pi$ = $12.902 \pm 0.045\ $s as the definitive value for the mean period spacing of NGC\,246. In Table~\ref{tab:avrg-sectors} we indicate the theoretical periods from the fitted mean period spacing (second column) and the corresponding residuals (third column). In the last column, we indicate the possible identification of the fitted periods with $\ell= 1$ modes, according to the period spacing previously inferred, although they may all be associated with $\ell= 2$, as already mentioned. In addition, we show the corresponding linear fit in the upper panel of Fig.~\ref{fig:linear-fit}, while the residuals for this case are displayed in the lower panel. It is worth noticing the presence of some minima in the distribution of the residuals, which can be attributed to the effects of mode trapping caused by the existence of internal chemical transition regions \citep{1994ApJ...427..415K}.

\begin{table}
\setlength{\tabcolsep}{5pt}
\renewcommand{\arraystretch}{1}
\centering
\caption{List of the 17 periods combined from Sectors 3 and 30 for NGC~246.}
\begin{tabular}{cccc}
\hline
\noalign{\smallskip}
 $\Pi^{\rm O}$ & $\Pi^{\rm T}$ & $\delta\Pi_{\rm fit}$ &  $\ell$ \\
\ [s] & [s]  & [s] & \\
\noalign{\smallskip}
\hline
\noalign{\smallskip}
$1460.1195^{(*)}$ & $1459.08$  & $ 1.0395$  & $1 $   \\
$1471.251^{(*)}$& $1471.9818  $  & $ -0.7308$ & $1 $   \\
$1503.417$& $ $            & $  $       & ?   \\
$1511.465^{(*)}$& $1510.6872  $  & $0.7778$   & $1$ \\
$1523.431^{(*)}$& $1523.589  $   & $-0.158 $  & $1$ \\
$1546.130^{(*)} $& $1549.3926 $   & $-3.2626 $  & $1$   \\
$1557.7885$& $ $           & $  $       & ?   \\
$1562.9085^{(*)}$& $1562.2944 $  & $0.6141 $  & $1$  \\
$1575.1750^{(*)}$& $1575.1962 $   & $-0.0212 $  & $1$   \\
$1587.5720^{(*)}$& $1588.098 $   & $-0.526 $  & $1$\\
$1601.810^{(*)}$& $1600.9998 $   & $0.8102 $  & $1$  \\
$1615.040^{(*)}$& $1613.9016 $   & $1.1384 $  & $1$  \\
$1627.960^{(*)}$& $1626.8034 $   & $1.1566 $  & $1$  \\
$1767.2075^{(*)}$& $1768.7232 $   & $-1.5157 $  & $1$  \\
$1774.620$& $ $           & $  $       & ? \\
$1805.9880^{(*)}$& $1807.4286 $   & $-1.4406 $  & $1$ \\
$1822.4565^{(*)}$& $1820.3304 $   & $2.1261 $  & $1$  \\

\noalign{\smallskip}
\hline
\noalign{\smallskip}
\end{tabular}
\tablefoot{The periods marked with an asterisk are used for the linear least-square fit of Fig~\ref{fig:linear-fit}. The $\ell$ identification is based on the period spacing extracted from the list of periods (see text for details).}
\label{tab:avrg-sectors}
\end{table}

\subsection{Mass determination from the comparison between the observed mean period spacing and the average of the computed period spacings}
\label{mass-period-spacing}

Now, we determine the mass of NGC\,246 by comparing the average of the computed period spacings, $\overline{\Delta  \Pi_k}$, for our grid of models, with the observed period spacing, $\Delta \Pi$, as determined in Sect.~\ref{period-spacing}. The average of the computed period spacings is calculated as $\overline{\Delta  \Pi_k}= (n-1)^{-1}  \sum_{k}  \Delta \Pi_k  $, where the    ``forward''    period spacing is defined as $\Delta \Pi_k= \Pi_{k+1}-\Pi_k$  ($k$ being the radial order) and $n$ is the number of theoretical periods within the range of the periods observed in the target star. For this star, $\Pi_k \in [1459,1823\ $s].

\begin{figure}  
\centering   \includegraphics[clip,width=260pt]{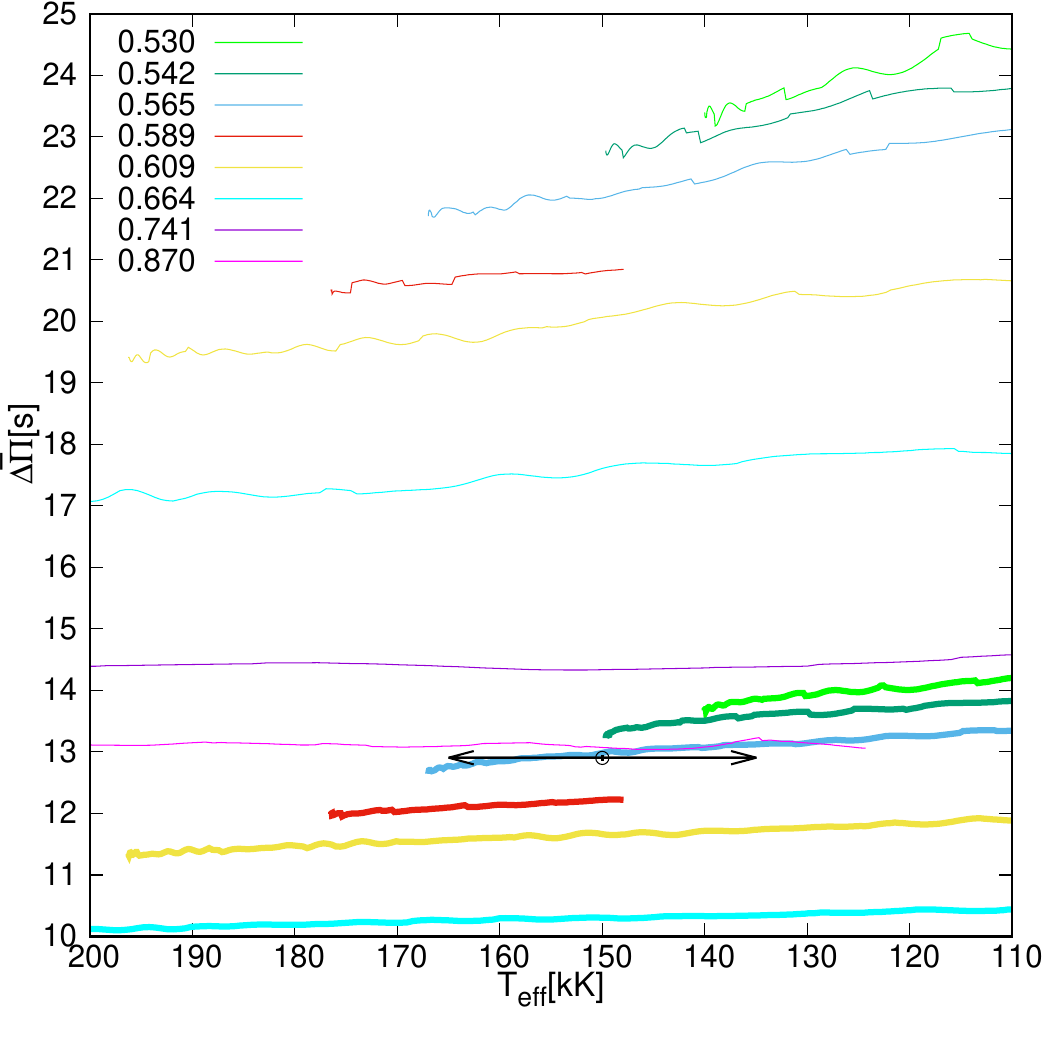} 
\caption{Average of the computed period spacings, $\overline{\Delta \Pi_k}$, assessed in the range of periods observed in NGC\,246, corresponding to our PG~1159 evolutionary sequences (with stellar masses expressed in solar units), in terms of the effective temperature. Thin curves represent the case in which the period spacing is assumed to be associated with $\ell= 1$ modes, while thick curves, with $\ell= 2$ modes. The location of NGC\,246 ($\Delta \Pi= 12.902 \pm 0.045\ $s, $T_{\rm eff}= 150\,000\pm 15\,000\ $K) is emphasized using a black circle with error bars.}
\label{fig:pspacing}
\end{figure}

In Fig.~\ref{fig:pspacing} we show the run of the average of the computed period spacings for NGC\,246 for the case in which the period spacing is assumed to be associated with $\ell= 1$ modes, indicated with thin curves, and $\ell= 2$ modes, represented with thick curves, in terms of $T_{\rm eff}$ for our PG 1159 evolutionary sequences. The location of NGC\,246 in this $\overline{\Delta  \Pi}-T_{\rm eff}$ plane is represented by a black circle with error bars, where we used the period spacing (and its uncertainty) determined in Sect.~\ref{period-spacing}, and the star's effective temperature (with a $10\%$ for its uncertainty, that is, slightly larger than the one given by the latest spectroscopic results).
 Clearly, for each set of curves with $\ell= 1$ and $2$, the lower the values of $\overline{\Delta  \Pi_k}$, the greater the stellar mass. When we consider the measured period spacing $\Delta \Pi= 12.902 \pm 0.045\ $s to be associated with $\ell= 1$ modes, it is clear that the stellar mass is slightly larger than $0.87\ M_{\sun}$. If we, instead, consider that the period spacing corresponds to $\ell= 2$ modes, a linear interpolation between $\Delta \Pi$ and $\overline{\Delta  \Pi_k}$\footnote{When there were no points to perform the linear interpolation, we extrapolated the theoretical values of $\overline{\Delta  \Pi_k}$.} yields a stellar mass of $M_{\star}= 0.568 ^{+0.006}_{-0.012} \ M_{\sun}$.

\begin{figure}  
\centering   \includegraphics[clip,width=260pt]{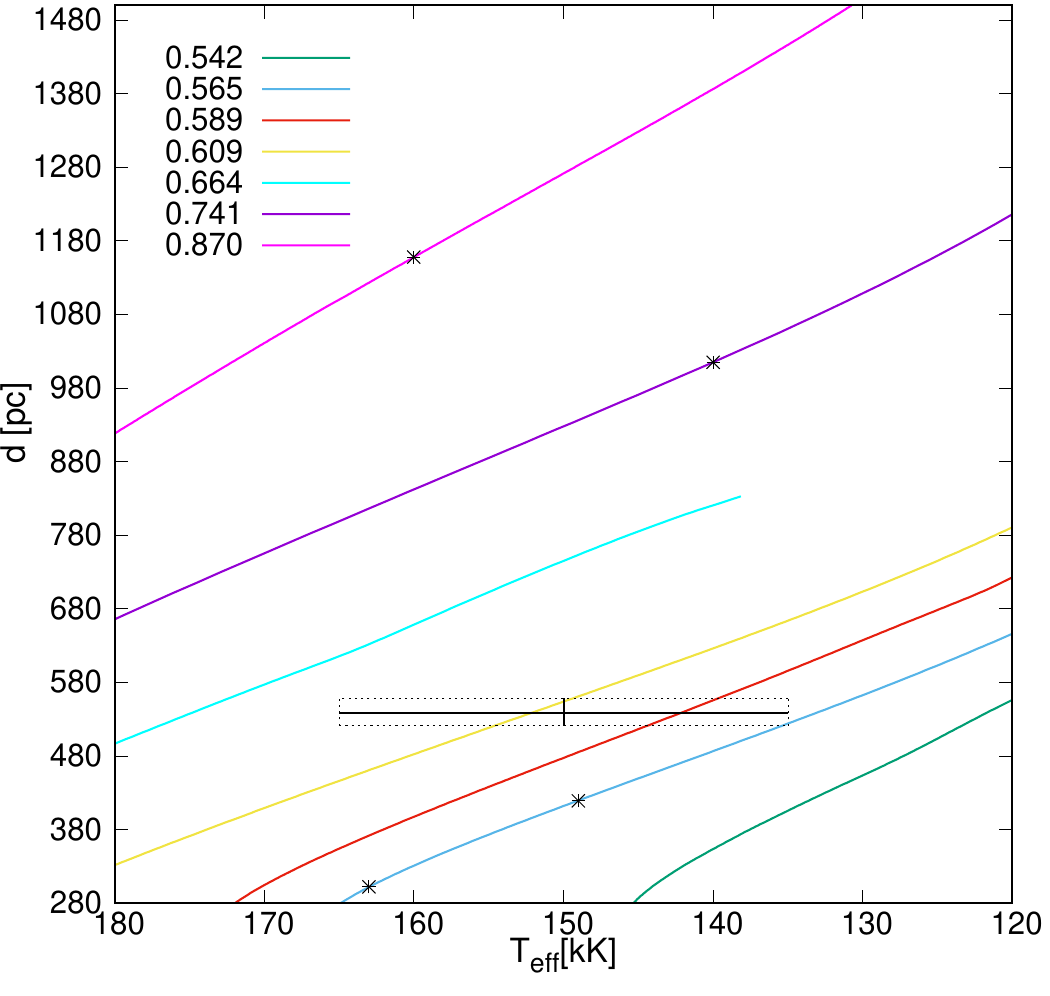} 
\caption{Inferred distance for our PG~1159 evolutionary models to match the observed apparent magnitude of NGC 246, in terms of effective temperature. The symbol with error bars marks the effective temperature reported from spectroscopy and the distance constraint from Gaia astrometry. The four individual models that achieve the best fits to the individual pulsation periods are marked with asterisks (see discussion in text).}
\label{fig:distance}
\end{figure}

PG 1159 stars with these two candidate seismic masses would have significantly different absolute magnitudes. We can compute the distances that the models would have to be at to match the apparent magnitude observed for NGC~246, following a procedure similar to previous works of this series \citep[see also][]{2019A&A...632A..42B,2023MNRAS.526.2846U}. 
The apparent visual magnitude of NGC~246, $m_V$, is $11.8\ $mag \citep{2012yCat.1322....0Z}, while the interstellar absorption $A_V(d)=  R_V\ E(B-V)$ (with $R_V = 3.1$) for this star is $0.062\ $mag, according to \cite{2016MNRAS.455.1459F} and \cite{2023Ap&SS.368...34A}. Using the bolometric correction extracted from the grids of He-atmosphere WDs of the Montreal Group\footnote{\url{https://www.astro.umontreal.ca/~bergeron/CoolingModels/}} \citep{Bergeron_1995,2006AJ....132.1221H,2020ApJ...901...93B}, the visual absolute magnitude is calculated as $M_V= M_B- BC$, where $M_B=  M_{B \sun} - 2.5 \times \log(L_{\star}/L_{\sun})$ and $ M_{B \sun}= 4.74$ \citep{2000asqu.book.....C}. 
In this way, the distance, $d$, can be inferred from $\log(d)= [m_V - M_V + 5 - A_V(d)]/5$. A plot of distances computed for our PG 1159 evolutionary models of different masses is shown in Figure~\ref{fig:distance}. Error bars indicate the distance estimate, $d_G= 538^{+20}_{-17}\ $pc \citep{2021AJ....161..147B}, and the spectroscopic effective temperature. Using other reported values of the interstellar absorption for this star, that is, $A_V= 0.14\ $mag \citep{2021A&A...656A..51G}, and $A_V= 0.0806\ $mag, that results from the value of $E(B-V)$ obtained from the 3D reddening map \citep{2014A&A...561A..91L,2017A&A...606A..65C,2018A&A...616A.132L}\footnote{\url{https://stilism.obspm.fr/}}, according to the galactic coordinates of this star, $(l,b)= (118.^{\circ}8630881437, -74.^{\circ}7090941475)$ does not significantly change the resulting distance estimations (the difference being within some tens of pc). While neither candidate mass suggested by the mean period spacing falls in the one-sigma error box, the lower-mass models are more compatible with the astrometric distance, leading us to favor the interpretation that we are observing an overtone sequence of $\ell=2$ modes and the corresponding $0.568 ^{+0.006}_{-0.012} \ M_{\sun}$ mass.

In closing this section, it is worth mentioning that these two candidate seismic masses would have different surface gravity as well: the one with $\sim 0.87\ M_{\sun}$ at $\sim 150\,000$, is characterized by $\log g \sim 5.52$, while the one with $0.568\ M_{\sun}$ at $\sim 150\,000$, by $\sim 6.29$. Considering that the value given by spectroscopy for NGC 246 is $5.7$, the former has a more similar value.

\subsection{Constraints from the individual observed periods:
searching for the best-fit model}
\label{period-to-period-fits}

We now turn to fitting the pulsation modes to stellar models to try to derive the stellar mass and compare the results with those obtained with the mean period spacing. This procedure, as already explained, may allow us to determine the mode identification of individual modes.
To carry this procedure out, we took different sets of periods into account: each set of periods from Sectors 3 and 30 from Table~\ref{F_list}, and the combination of both, as indicated in Table~\ref{tab:avrg-sectors}. In this last case, we additionally considered two particular subsets of the combination: one subset with 11 periods  
(see Sect.~\ref{rotational}), and another one with the 14 periods marked with asterisks in Table~\ref{tab:avrg-sectors}, that fit the mean period spacing determined in Sect.~\ref{period-spacing}. For all cases, we first considered that all the periods are associated with $\ell= 1$ $g$ modes and employed them to assess the quality function given by Eq. (\ref{chi}). Next, we repeated the procedure, but considering that periods result from a mixture of $\ell= 1$ and $\ell= 2$ $g$ modes. Generally, if a single maximum exists, we can adopt the corresponding model as the asteroseismic solution. Unfortunately, in the cases we studied here there are multiple possible solutions, and then we need to apply an external constraint in order to adopt a single asteroseismic model. In this case, the constraint is the effective temperature and its uncertainty (where, again, we took a more flexible value than the one given by spectroscopy).

\begin{figure}  
\centering   \includegraphics[clip,width=250pt]{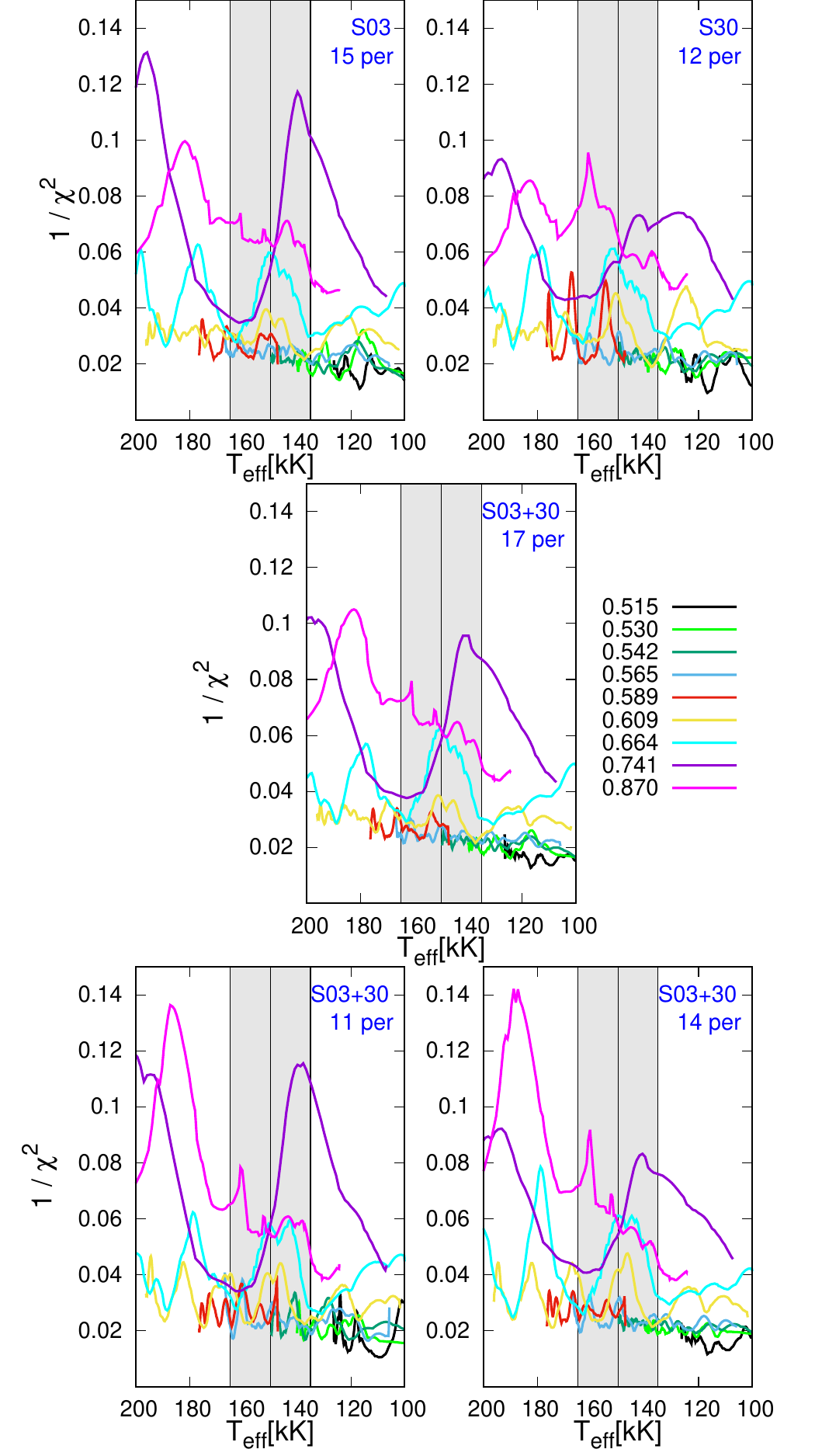}
\caption{Inverse of the quality function, $1/\chi^2$, versus $T_{\rm eff}$ of the period-to-period fits in the $\ell= 1$ case, for the five sets of periods considered for NGC~246: Sector 3 (\textit{upper left panel}), Sector 30 (\textit{upper right panel}), the 17 periods from the combination of both sectors (\textit{middle panel}), and the two subsets of the combination of both sectors with 11 (\textit{lower left panel}) and 14 periods (\textit{lower right panel}). The vertical gray strip indicates the spectroscopic values of $T_{\rm eff}$ and the corresponding uncertainty for this star.}
\label{fig:l1-all}
\end{figure}

We start by considering the list of 15 periods from Sector 3 (Table~\ref{F_list}), and we show the results for the $\ell= 1$ case in the upper left panel of Fig.~\ref{fig:l1-all}. In the figure, we plot the inverse of the quality function, $1/\chi^2$, such that the greater the value of $1/\chi^2$, the better the fit between theoretical and observed periods. It is clear that there is more than one possible asteroseismic solution. In particular, the absolute maximum (the best solution) in this case corresponds to a model with $T_{\rm eff} \sim 196\,000\ $K, 
which is much higher than the values of $T_{\rm eff}$ allowed by the spectroscopic determinations for NGC\,246, and can be discarded. The second-best global fit lies within the constraints given by the allowed range of effective temperatures at $\sim 140\,000\ $K, for $M_{\star}= 0.741\ M_{\sun}$ and $1/\chi^2= 0.117$. 
When we consider the list of 12 periods from Sector 30 (Table~\ref{F_list}), and repeat the above procedure, we obtain the results shown in the upper right panel of Fig.~\ref{fig:l1-all}. Once again, there are multiple possible solutions, but the best-global fit lies within the range of allowed $T_{\rm eff}$ at $\sim 160\,000\ $K, for $0.87\ M_{\sun}$ and $1/\chi^2= 0.096$. 
Next, we consider the set of 17 periods combining Sectors 3 and 30, as listed in Table~\ref{tab:avrg-sectors}. The results are shown in the middle panel of Fig.~\ref{fig:l1-all}. The best-global fit lies at a very high $T_{\rm eff}$, and within the ranges of allowed $T_{\rm eff}$, there is a possible solution characterized by $\sim 140\,000\ $K, for $0.741\ M_{\sun}$ and $1/\chi^2= 0.097$. 
The lower left panel of Fig.~\ref{fig:l1-all} depicts the results when considering the subset of 11 periods. In this case, the best-global fit lies at high values of $T_{\rm eff}$. However, in the allowed ranges of $T_{\rm eff}$, there is a possible solution characterized by $\sim 140\,000\ $K, for $0.741\ M_{\sun}$ and $1/\chi^2= 0.116$.
The results for the subset 14 periods are shown in the lower right panel of Fig.~\ref{fig:l1-all}. Similarly to the previous case, although the best-global fits lie at very high values of $T_{\rm eff}$, there are some possible solutions, with comparable quality, given by $\sim 160\,000\ $K, for $0.870\ M_{\sun}$ and $1/\chi^2= 0.092$ and $\sim 140\,000\ $K, for $0.741\ M_{\sun}$ and $1/\chi^2= 0.083$. It is clear from the figure that the S03 15 periods (first panel) results are similar to the S03+30 11 periods (lower left panel) results.

\begin{figure}  
\centering   \includegraphics[clip,width=250pt]{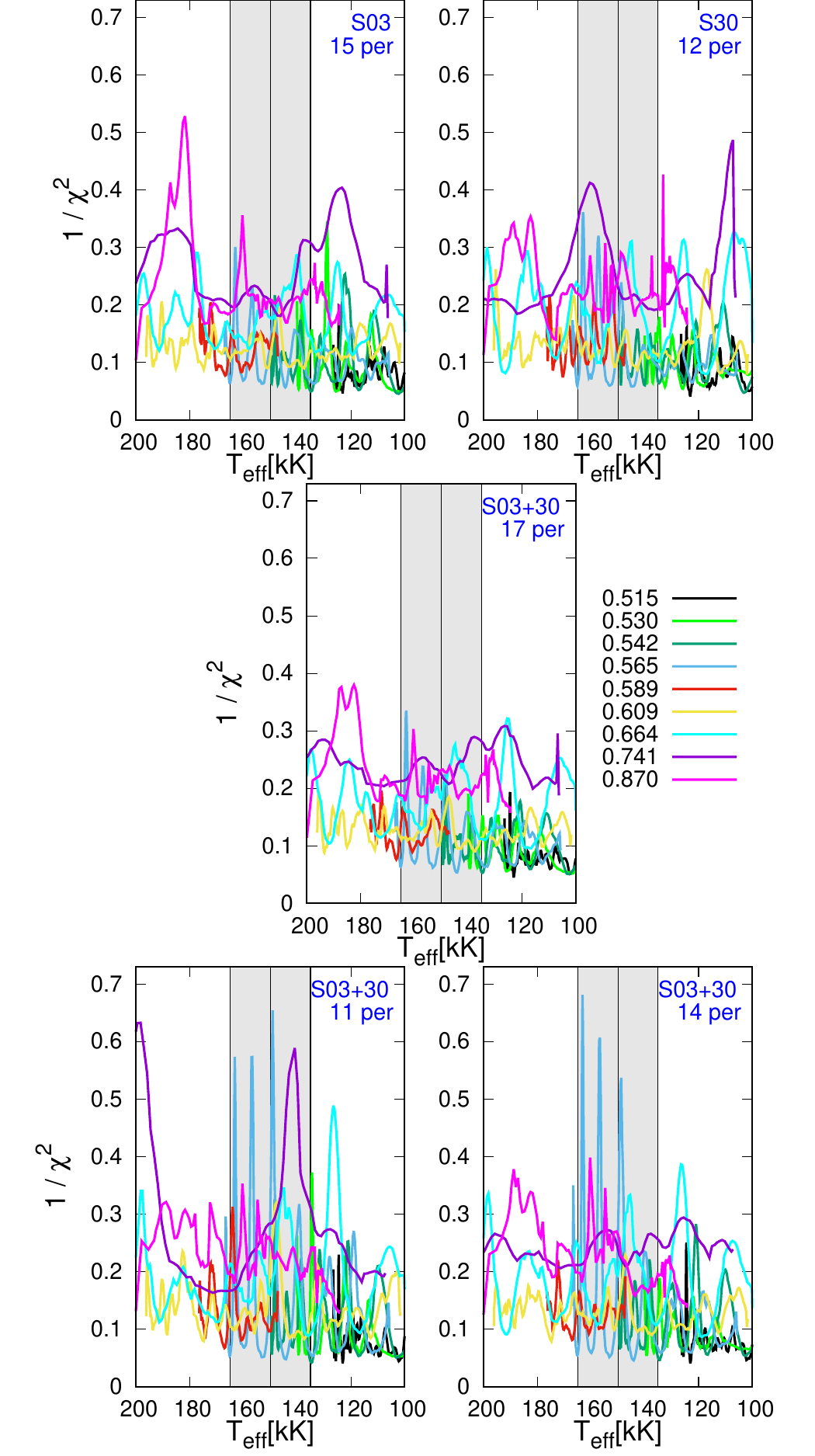}
\caption{Same as Fig.~\ref{fig:l1-all}, but for the $\ell= 1,2$ case. Clearly, the $1/\chi^2$ values are much higher than for the $\ell= 1$ case (Fig.~\ref{fig:l1-all}), indicating these models fit the data better. The middle panel demonstrates that there is not a particularly preferred model fit in the case of all 17 periods from the combined sectors. Regarding the two bottom panels, the $0.565\ M_{\sun}$ sequence has several models that fit the data well within the observed temperature bounds. In the 11-period case, there is also a $0.74\ M_{\sun}$ model that fits.}
\label{fig:l1l2-all}
\end{figure}

Regarding the $\ell= 1, 2$ case, we show the results in Fig.~\ref{fig:l1l2-all}. Clearly, the quality of the solutions is in general improved, as evidenced by the greater values of  $1/\chi^2$. The results for the set of periods from Sector 3 are shown in the upper left panel. The best global fits are located at higher and lower values of $T_{\rm eff}$ than those given by spectroscopy. However, there are possible solutions within the allowed ranges of $T_{\rm eff}$ at $\sim 160\,000\ $K, for $0.870\ M_{\sun}$ and $1/\chi^2= 0.356$, at $\sim 140\,000\ $K, for $0.741\ M_{\sun}$ and $1/\chi^2= 0.313$, and also at $\sim 163\,000\ $K, for $0.565\ M_{\sun}$ and $1/\chi^2= 0.301$.
For Sector 30, as shown in the upper right panel of Fig.~\ref{fig:l1l2-all}, the best solution has a very low $T_{\rm eff}$, and the solutions that lie within the range of allowed $T_{\rm eff}$ are characterized by $\sim 160\,000\ $K, $0.741\ M_{\sun}$, and $1/\chi^2= 0.412$, and $\sim 163\,000\ $K, $0.565\ M_{\sun}$, and $1/\chi^2= 0.361$.
For the combined set of periods from both sectors, as the middle panel of Fig.~\ref{fig:l1l2-all} depicts, the best-fit models have very high values of $T_{\rm eff}$. Still, there are three possible solutions within the range of allowed $T_{\rm eff}$ that are worth mentioning at $\sim 163\,000\ $K, for $0.565\ M_{\sun}$ and $1/\chi^2= 0.335$, at $\sim 160\,000\ $K, for $0.870\ M_{\sun}$ and $1/\chi^2= 0.303$, and also at $\sim 140\,000\ $K, for $0.741\ M_{\sun}$ and $1/\chi^2= 0.289$.
The case of the subset with 11 periods, represented in the lower left panel of Fig.~\ref{fig:l1l2-all}, shows the best-global fit for a model characterized by $\sim 150\,000\ $K, $0.565\ M_{\sun}$ and $1/\chi^2= 0.654$. Other possible solutions within the range of allowed $T_{\rm eff}$ are characterized by $\sim 140\,000\ $K, $0.741\ M_{\sun}$ and $1/\chi^2= 0.589$, $\sim 157\,000\ $K, $0.565\ M_{\sun}$ and $1/\chi^2= 0.573$, and $\sim 163\,000\ $K, $0.565\ M_{\sun}$ and $1/\chi^2= 0.574$.
Finally, the lower right panel of Fig.~\ref{fig:l1l2-all} depicts the results for the subset with 14 periods. In this case, the best-global fit is represented by a model with $\sim 163\,000\ $K,  $0.565\ M_{\sun}$ and $1/\chi^2= 0.681$. It is worth mentioning that this solution has the best quality (the highest value of $1/\chi^2$) among all the cases considered in this work. Other possible solutions, that also lie within the range of allowed $T_{\rm eff}$, are characterized by models with $\sim 157\,000\ $K, $0.565\ M_{\sun}$ and $1/\chi^2= 0.607$, and $\sim 150\,000\ $K, $0.565\ M_{\sun}$ and $1/\chi^2= 0.537$.

When we compare the results from all cases, it is clear that the global behavior of the quality function changes, although some solutions are repeated in more than one case. 
 Within the allowed ranges of $T_{\rm eff}$, four solutions stand out. When using only $\ell= 1$ modes, and also when using a mixture of $\ell= 1, 2$ modes, there are possible seismic solutions characterized by $0.741\ M_{\sun}$ and $\sim 140\,000\ $K, and $0.870\ M_{\sun}$ and $\sim 160\,000\ $K. On the other hand, if we consider $\ell= 1, 2$ modes,  there are also possible seismic solutions with $0.565\ M_{\sun}$ and $\sim 163\,000\ $K, and $0.565\ M_{\sun}$ and $\sim 150\,000\ $K.  

\begin{figure}  
\centering   \includegraphics[clip,width=260pt]{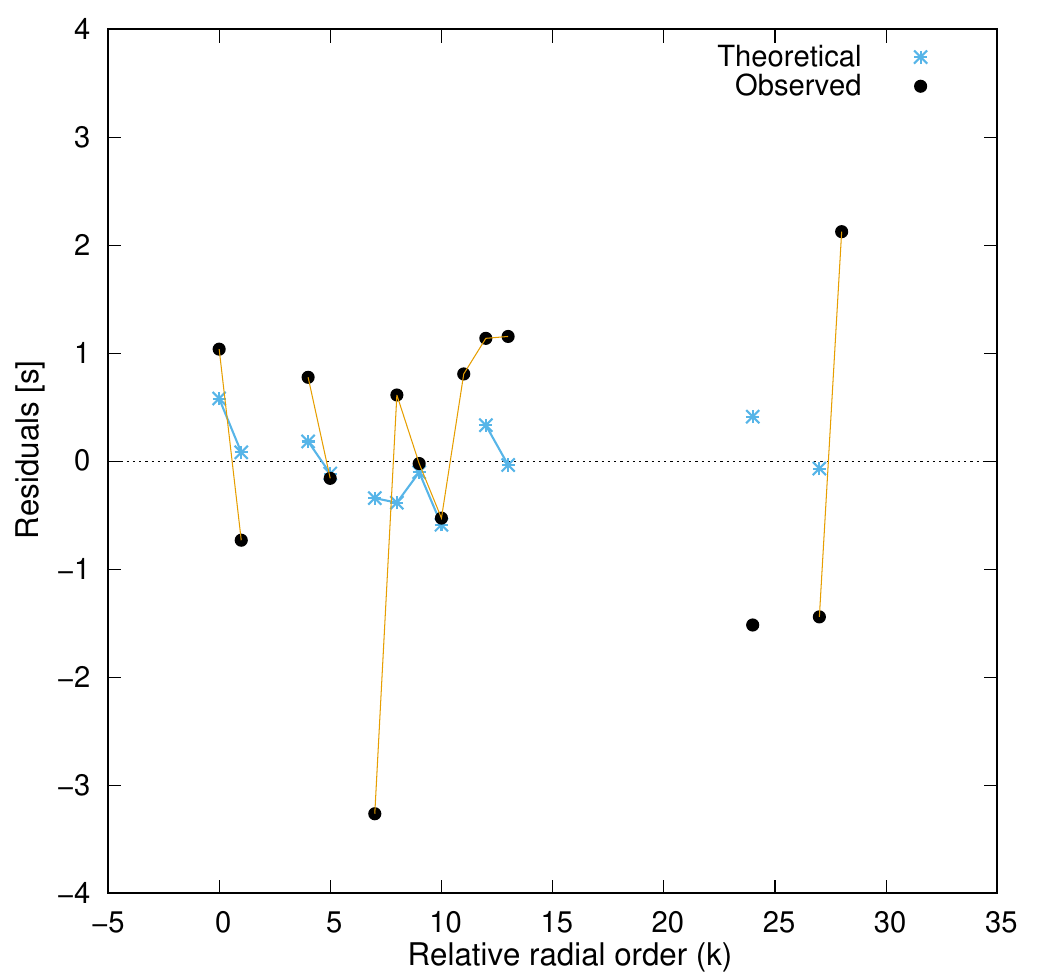} 
\caption{Distribution of the residuals relative to the mean period spacing for the case of the observed periods (black points, see bottom panel of Fig.~\ref{fig:linear-fit}), along with the case for the theoretical periods (sky-blue crosses) of the model fit with $0.565\ M_{\sun}$ and $\sim 163\,000\ $K, represented in terms of the relative radial order. As in Fig.~\ref{fig:linear-fit}, thin lines connect modes with consecutive radial order.}
\label{fig:obs-model-residuals}
\end{figure}

Although it is not possible to adopt one solution, it is of interest to study if there is agreement between the data residuals and the model residuals of an asteroseismic solution, instead of only comparing the individual periods. We have done so with our possible asteroseismic solutions, but we only show the case of the model with $0.565\ M_{\sun}$ and $\sim 163\,000\ $K in Fig~\ref{fig:obs-model-residuals}, given that it has the largest $1/\chi^2$ value for all the cases considered in this work. In the figure, we plot the residuals of the period distribution relative to the observed period spacing (black points, as in the bottom panel of Fig.~\ref{fig:linear-fit}), in terms of the relative radial order, along with the corresponding residuals for the mentioned theoretical model. As can be seen, there is a moderate agreement between the two.

Given these four best-fitting models, we calculate their asteroseismic distances and compare them with the distance resulting from the {\it Gaia} measurements. The masses, temperatures, luminosities, and inferred distances for these models are listed in Table~\ref{tab:fourmodels}, and these models are marked with asterisks in Figure~\ref{fig:distance}.
It is clear that none of these four candidate models falls within the uncertainties of the precise distance constraint from \textit{Gaia}. 

\begin{table*}
\setlength{\tabcolsep}{5pt}
\renewcommand{\arraystretch}{1}
\centering
\caption{Properties and inferred distances for four best-fitting seismic models. The sixth column indicates which models fit well to the data when just considering $\ell=1$ modes or a mixture of $\ell=1, 2$ modes. See text for discussion of how these models were identified.  Also shown are the corresponding values given by spectroscopy and astrometry, for quick comparison.}
\begin{tabular}{ccccccc}
\hline
\noalign{\smallskip}
 $M_{\star}$ & $T_{\rm eff}$ & $\log g$ & $\log(L_{\star}/L_{\sun}) $ &  distance & modes used  & Reference\\
\ $[M_{\sun}]$ & $[K]$ & [cgs] & & [pc] &  $\ell$  &\\
\noalign{\smallskip}
\hline
\noalign{\smallskip}
0.741 & 140\,000 & 5.59 & 4.25 & 1016 & $1$ and $1$,$2$ & This work\\ %ell=1
0.870 & 160\,000 & 5.64 & 4.51 & 1158 & $1$ and $1$,$2$  & This work\\ %ell=1
0.565 & 163\,000 & 6.62 & 3.37 & 302   & $1$,$2$ & This work\\ %ell=1 or 2
0.565 & 150\,000 & 6.29 & 3.55 & 419   & $1$,$2$ & This work\\ %ell=1 or 2
\noalign{\smallskip}
\hline
\noalign{\smallskip}

 0.74   & 150\,000   & 5.7 & 4.27   & 970 & & Spectroscopy \\
\noalign{\smallskip}
\hline
\noalign{\smallskip}
    & &  & & 538  & & Astrometry\\
\noalign{\smallskip}
\hline
\noalign{\smallskip}    
\end{tabular}
\label{tab:fourmodels}
\end{table*}

Finally, a calculation of the stellar mass when fixing the distance of NGC\,246 to the value given by {\sl Gaia} for different values of $A_V$ and some effective temperatures in the range of interest, results in model sequences characterized by a stellar mass around $ 0.6\ M_{\sun}$. For the resulting luminosities to be compatible with stellar masses $\gtrsim 0.7\  M_{\sun}$, $A_V$ should be $\gtrsim 1.3\ $mag, which, then, would be too large and inconsistent with the high value of the galactic latitude of NGC\,246 below the Galactic plane ($b= -74.^{\circ}7090941475$).

\section{Summary and discussion}  
\label{conclusions}

In this paper, we have presented a detailed asteroseismic study of NGC~246, a pulsating PG~1159 star, based on the high-precision photometry data from {\it TESS} observations. NGC~246 (\Teff $\sim 150\,000\ $K and $\logg \sim 5.7$) was observed by {\it TESS} in Sector 3, at the short $120\ $s cadence, and 30, at the short $120\ $s and ultra-short $20\ $s cadence. Our frequency analysis revealed a total of 17 periodicities. These oscillation periods ($1460$ - $1823\ $s) are associated with nonradial $g$ modes. Given this frequency spectrum, we investigated the presence of rotation multiplets. However, we did not find any distinct indications of rotational splitting for either dipole or quadrupole modes.

Employing the inferred pulsation periods, we carried out an asteroseismic analysis on NGC~246 by means of our fully evolutionary models of PG~1159 stars. We first found a constant period spacing underlying the observed periods, which was used to make inferences on the stellar mass via the comparison between the observed mean period spacing and the average of the computed period spacings. Our results indicate that if the observed mean period spacing is associated with $\ell= 1$ $g$ modes, then $M_{\star} \gtrsim 0.87\ M_{\sun}$. If, instead, it is associated with $\ell= 2$ $g$ modes, then $M_{\star} \gtrsim 0.568\ M_{\sun}$. Next, we searched for the best-fit model between the theoretical and the observed periods. For this procedure, we first assumed that all the periods were associated with $\ell= 1$ $g$ modes only, and subsequently, with a mixture of $\ell= 1, 2$ $g$ modes.
Although we did not find a clear and unambiguous solution for NGC~246, there are some possible solutions that lie within the values of effective temperature (for which we took  an error of $\pm 10\%$). When considering that all modes are $\ell= 1$ only, our results suggest a high-stellar mass ($\gtrsim 0.74\ M_{\sun}$), but when we allow the modes to take a mixture of $\ell= 1$ and $2$, both high ($\gtrsim 0.74\ M_{\sun}$) and intermediate ($\sim 0.57\ M_{\sun}$) stellar masses also fit the data well. It is worth mentioning that the $\ell =1$ fits have smaller $1/\chi^2$ values than the $\ell= 1, 2$ fits.
A comparison between the corresponding asteroseismic distances for these models to the {\sl Gaia} astrometric distance \citep[$538^{+20}_{-17}$\,pc,][]{2021AJ....161..147B} indicates that only one solution is in agreement, which corresponds to a model with $M_{\star}= 0.565\ M_{\sun}$, lying at approximately the same \Teff of the star.

 NGC~246 has been recently analyzed by \cite{Loebling2018,Loebling2020} who, employing new spectroscopic observations and state-of-the-art non-LTE models, found \Teff = $150\,000 \pm 10\,000\ $K and $\logg = 5.7 \pm 0.1$. This results in $M_{\star}= 0.74^{+0.19}_{-0.23}$\Msol\, and then, in a stellar radius of $R_{\star}= 0.20$\,\Rsol. Interestingly, in the spectral analysis carried out by \cite{RauchWerner1997}, the authors noted unusually broad spectral line cores, which was interpreted as a signature of stellar rotation with velocity $v \sin i \approx 70$~km\,s$^{-1}$, being $i$ the unknown inclination angle of the rotation axis. This value was later confirmed by using high-resolution spectra by \cite{RauchKoeper2004}. More recently, \cite{Loebling2018} measured $v \sin i = 75\pm15$~km\,s$^{-1}$. Considering the latter along with the stellar radius derived from spectroscopy, it would imply a rotational period of $P\times\sin i = 0.68\pm0.14$~d. If this were the rotational period that characterizes NGC~246, a rotational splitting of $8.51\ \mu$Hz  would be expected in the frequency spectrum in $\ell=1$ modes.  However, neither the amplitude spectra of Sectors 3 and 30 exhibit a clear pattern that would confirm this expectation. Still, it must be noted that the rotational velocity could be overestimated because, e.g., microturbulent motions in the atmosphere could contribute to the line-core broadening. Additionally,  \citet{Loebling2020} determined a spectroscopic distance of $970\pm200$\,pc for NGC~246. This is a factor of 1.7 larger than the {\tt Gaia} distance. In order to bring the spectroscopic distance in agreement with the parallax distance, an increase of the surface gravity by a factor of three (that is, an increase of 0.5 dex in \logg) would be necessary. As demonstrated by \citet{Loebling2020}, such a high value for the gravity strongly contradicts the model fits to the \ion{He}{ii} line profiles. The reason for this problem remains unknown. Interestingly, our possible asteroseismic solution with $0.565\ M_{\sun}$ and $\sim 150\,000\ $K, has a value of \logg (= 6.29), which is in line with the value needed to fit the parallax distance. 

Besides the distance resulting from the {\sl Gaia} parallax value, two independent estimations for the distance to NGC~246 were performed in the past. First, by fitting the resolved companion to the zero-age main sequence, \cite{BondCiardullo1999} found a distance of $495^{+145}_{-100}$~pc, which is in agreement with the parallax distance. Second, from the observed angular expansion rate of the PN \citep{Liller1966} and a measurement of the expansion velocity, \cite{Terzian1997} found a distance of 570~pc. However, the errors in the expansion rate and velocity must be regarded as too high to favor either the parallax or the spectroscopic distance. Modern precise determinations of expansion rates of three PNe in a distance of about 2~kpc using HST imaging over four years were successfully performed by \cite{Palen2002}. In principle, such measurements are also feasible for NGC~246 and could be compared to the precise astrometry measurements.

 All in all, our results show two possible outcomes in the derivation of the stellar mass. The use of the pulsation periods via the comparison between the observed period spacing and the average of the computed period spacings points to a very high-stellar mass asteroseismic model if the derived mean period spacing is associated with $\ell= 1$ $g$ modes, and an intermediate stellar mass if, instead, it is associated with $\ell= 2$ $g$ modes. Although we were not able to adopt an asteroseismic model from the period-to-period fits, the results obtained from the different sets of periods considered in this procedure would indicate high-stellar mass models for the $\ell= 1$ case, and both high- and intermediate-stellar mass models for the $\ell= 1, 2$ case. Astrometry, by means of the precise measurements from Gaia, seems to point toward intermediate stellar masses, making one of our possible asteroseismic solutions particularly promising. At the same time, abiding by the precise distance from Gaia, our asteroseismological results may indicate that the set of modes of this star should be a mixture of $\ell = 1, 2$. Regarding the stellar rotation, the conclusions derived from the spectroscopic analysis are not consistent with the precise astrometric distance, an aspect that cannot be ignored. Concurrently, asteroseismology does not so far provide a conclusive value for the rotation period.

 Our lack of conclusive results shows the enigmatic nature of NGC\,246. At this point, isolating the specific reason(s) behind the disparities among spectroscopic, seismological, and astrometric ({\sl Gaia}) masses remains elusive, and clearly, further work is required. Hopefully, future follow-up observations from space, such as the PLAnetary Transits and Oscillations mission \citep[PLATO,][]{2018EPSC...12..969P}, or ground-based program, such as BlackGEM \citep{2015ASPC..496..254B}, might help in shedding some light on this open problem. Additionally, improved atmospheric parameters from high-resolution spectroscopic data are necessary to put better constraints on the modeling and to, possibly, help mitigate the discrepancies found in this work.
  
\begin{acknowledgements}
We want to thank our anonymous referee for the comments and suggestions that greatly improved the original version of the paper.

Part of this work was supported by AGENCIA through the Programa de Modernizaci{\'o}n Tecnol{\'o}gica BID 1728/OC-AR, and by the PIP 112-200801-00940 grant from CONICET.
%:
This research was supported in part by the National Science Foundation under Grant No. NSF PHY-1748958, which enabled the KITP Program on White Dwarfs as Probes of the Evolution of Planets, Stars, the Milky Way and the Expanding Universe in 2022, at which some of this work was performed.
M. U. gratefully acknowledges funding from the Research Foundation Flanders (FWO) by means of a junior postdoctoral fellowship (grant agreement No. 1247624N). PS and GH thank the Polish National Center for Science (NCN) for supporting the study through grants 2015/18/A/ST9/00578 and 2021/43/B/ST9/02972. 

This paper includes data collected with the {\it TESS} mission, obtained from the MAST data archive at the Space Telescope Science Institute (STScI). Funding for the {\it TESS} mission is provided by the NASA Explorer Program. This work has made use of data from the European Space Agency (ESA) mission Gaia (\url{https://www.cosmos.esa.int/gaia}), processed by the {\it Gaia} Data Processing and Analysis Consortium (DPAC, \url{https://www.cosmos.esa.int/web/gaia/dpac/consortium}). Funding for the DPAC has been provided by national institutions, in particular, the institutions participating in the Gaia Multilateral Agreement. This research has made use of NASA's Astrophysics Data System Bibliographic Services, and the SIMBAD and VizieR databases, operated at CDS, Strasbourg, France.

\end{acknowledgements}

\bibliographystyle{aa}
\bibliography{biblio}

\end{document}